\documentclass[11pt]{article}
\pdfoutput=1
\usepackage{hyperref}
\usepackage{graphicx}
\usepackage{epsfig}
\usepackage{epsf}
\usepackage{epstopdf}

\begin{document}
\pagestyle{empty}
\def\eqa{\!\!&=&\!\!}
\def\ccr{\nonumber\\}

\def\la{\langle}
\def\ra{\rangle}

\def\del{\Delta}
\def\ddel{{}^\bullet\! \Delta}
\def\deld{\Delta^{\hskip -.5mm \bullet}}
\def\ddeld{{}^{\bullet}\! \Delta^{\hskip -.5mm \bullet}}
\def\dddel{{}^{\bullet \bullet} \! \Delta}

\newcommand{\ba}{\begin{array}}
\newcommand{\ea}{\end{array}}
\newcommand{\identy}{1\!\!1}
\def\ni{\noindent}
\def\la{\langle}
\def\ra{\rangle}

\def\rld{\rlap{\,/}D}
\def\rldd{\rlap{\,/}\nabla}
%
\def\half{{1\over 2}}
\def\third{{1\over3}}
\def\fourth{{1\over4}}
\def\fifth{{1\over5}}
\def\sixth{{1\over6}}
\def\seventh{{1\over7}}
\def\eigth{{1\over8}}
\def\ninth{{1\over9}}
\def\tenth{{1\over10}}
\def\bN{\mathop{\bf N}}
\def\R{{\rm I\!R}}
\def\Eins{{\mathchoice {\rm 1\mskip-4mu l} {\rm 1\mskip-4mu l}
{\rm 1\mskip-4.5mu l} {\rm 1\mskip-5mu l}}}
\def\Z{{\mathchoice {\hbox{$\sf\textstyle Z\kern-0.4em Z$}}
{\hbox{$\sf\textstyle Z\kern-0.4em Z$}}
{\hbox{$\sf\scriptstyle Z\kern-0.3em Z$}}
{\hbox{$\sf\scriptscriptstyle Z\kern-0.2em Z$}}}}
\def\abs#1{\left| #1\right|}
\def\com#1#2{
        \left[#1, #2\right]}
\def\square{\kern1pt\vbox{\hrule height 1.2pt\hbox{\vrule width 1.2pt
   \hskip 3pt\vbox{\vskip 6pt}\hskip 3pt\vrule width 0.6pt}
   \hrule height 0.6pt}\kern1pt}
      \def\boxop{{\raise-.25ex\hbox{\square}}}
\def\contract{\makebox[1.2em][c]{
        \mbox{\rule{.6em}{.01truein}\rule{.01truein}{.6em}}}}
\def\ltap{\ \raisebox{-.4ex}{\rlap{$\sim$}} \raisebox{.4ex}{$<$}\ }
\def\gtap{\ \raisebox{-.4ex}{\rlap{$\sim$}} \raisebox{.4ex}{$>$}\ }
\def\mn{{\mu\nu}}
\def\rs{{\rho\sigma}}
\newcommand{\Det}{{\rm Det}}
\def\Tr{{\rm Tr}\,}
\def\tr{{\rm tr}\,}
\def\sumij{\sum_{i<j}}
\def\e{\,{\rm e}}
\def\pa{\partial}
\def\dA{\partial^2}
\def\ddx{{d\over dx}}
\def\ddt{{d\over dt}}
\def\der#1#2{{d #1\over d#2}}
\def\lie{\hbox{\it \$}} 
\def\partder#1#2{{\partial #1\over\partial #2}}
\def\secder#1#2#3{{\partial^2 #1\over\partial #2 \partial #3}}
%
\newcommand{\be}{\begin{equation}}
\newcommand{\ee}{\end{equation}\noindent}
\newcommand{\bear}{\begin{eqnarray}}
\newcommand{\ear}{\end{eqnarray}\noindent}
\newcommand{\benn}{\begin{enumerate}}
\newcommand{\enn}{\end{enumerate}}
\newcommand{\veject}{\vfill\eject}
\newcommand{\ven}{\vfill\eject\noindent}
%
\def\eq#1{{eq. (\ref{#1})}}
\def\eqs#1#2{{eqs. (\ref{#1}) -- (\ref{#2})}}
%
\def\totint{\int_{-\infty}^{\infty}}
\def\posint{\int_0^{\infty}}
\def\negint{\int_{-\infty}^0}
\def\pint{{\dps\int}{dp_i\over {(2\pi)}^d}}
%
\newcommand{\GeV}{\mbox{GeV}}
\def\FFdual{F\cdot\tilde F}
\def\bra#1{\langle #1 |}
\def\ket#1{| #1 \rangle}
\def\braket#1#2{\langle {#1} \mid {#2} \rangle}
\def\vev#1{\langle #1 \rangle}
\def\rightvac{\mid 0\rangle}
\def\leftvac{\langle 0\mid}
\def\ihbar{{i\over\hbar}}
\def\slash#1{#1\!\!\!\raise.15ex\hbox {/}}
\newcommand{\slD}{\,\raise.15ex\hbox{$/$}\kern-.27em\hbox{$\!\!\!D$}}
\newcommand{\slpartial}{\raise.15ex\hbox{$/$}\kern-.57em\hbox{$\partial$}}
\newcommand{\cL}{\cal L}
\newcommand{\D}{\cal D}
\newcommand{\Dhalf}{{D\over 2}}
\def\eps{\epsilon}
\def\epshalf{{\epsilon\over 2}}
\def\lag{( -\partial^2 + V)}
\def\freeexp{{\rm e}^{-\int_0^Td\tau {1\over 4}\dot x^2}}
\def\kinb{{1\over 4}\dot x^2}
\def\kinf{{1\over 2}\psi\dot\psi}
\def\expk{{\rm exp}\biggl[\,\sum_{i<j=1}^4 G_{Bij}k_i\cdot k_j\biggr]}
\def\expp{{\rm exp}\biggl[\,\sum_{i<j=1}^4 G_{Bij}p_i\cdot p_j\biggr]}
\def\expshort{{\e}^{\half G_{Bij}k_i\cdot k_j}}
\def\expabb{{\e}^{(\cdot )}}
\def\epseps#1#2{\varepsilon_{#1}\cdot \varepsilon_{#2}}
\def\epsk#1#2{\varepsilon_{#1}\cdot k_{#2}}
\def\kk#1#2{k_{#1}\cdot k_{#2}}
\def\G#1#2{G_{B#1#2}}
\def\Gp#1#2{{\dot G_{B#1#2}}}
\def\GF#1#2{G_{F#1#2}}
\def\Dab{{(x_a-x_b)}}
\def\Dsq{{({(x_a-x_b)}^2)}}
\def\PITD{{(4\pi T)}^{-{D\over 2}}}
\def\4piTD{{(4\pi T)}^{-{D\over 2}}}
\def\4piT4{{(4\pi T)}^{-2}}
\def\TintmD{{\dps\int_{0}^{\infty}}{dT\over T}\,e^{-m^2T}
    {(4\pi T)}^{-{D\over 2}}}
\def\Tintm4{{\dps\int_{0}^{\infty}}{dT\over T}\,e^{-m^2T}
    {(4\pi T)}^{-2}}
\def\Tintm{{\dps\int_{0}^{\infty}}{dT\over T}\,e^{-m^2T}}
\def\Tint{{\dps\int_{0}^{\infty}}{dT\over T}}
\def\np{n_{+}}
\def\nm{n_{-}}
\def\Np{N_{+}}
\def\Nm{N_{-}}
\newcommand{\slG}{{{\dot G}\!\!\!\! \raise.15ex\hbox {/}}}
\newcommand{\Gd}{{\dot G}}
\newcommand{\Gund}{{\underline{\dot G}}}
\newcommand{\Gdd}{{\ddot G}}
\def\GBd12{{\dot G}_{B12}}
\def\Dx{\dps\int{\cal D}x}
\def\Dy{\dps\int{\cal D}y}
\def\Dpsi{\dps\int{\cal D}\psi}
\def\dint#1{\int\!\!\!\!\!\int\limits_{\!\!#1}}
\def\ddtau{{d\over d\tau}}
\def\ie{\hbox{$\textstyle{\int_1}$}}
\def\iz{\hbox{$\textstyle{\int_2}$}}
\def\id{\hbox{$\textstyle{\int_3}$}}
\def\ldop{\hbox{$\lbrace\mskip -4.5mu\mid$}}
\def\rdop{\hbox{$\mid\mskip -4.3mu\rbrace$}}
%
\newcommand{\1}{{\'\i}}
\newcommand{\no}{\noindent}
\def\non{\nonumber}
\def\dps{\displaystyle}
\def\sy{\scriptscriptstyle}
\def\sy{\scriptscriptstyle}

%

\newcommand{\bea}{\begin{eqnarray}}  
\newcommand{\eea}{\end{eqnarray}}  
\def\eqa{&=&}  
\def\ccr{\nonumber\\}  
  
\def\a{\alpha}
\def\b{\beta}
\def\m{\mu}
\def\n{\nu}
\def\r{\rho}
\def\s{\sigma}
\def\ep{\epsilon}

\def\cosech{\rm cosech}
\def\sech{\rm sech}
\def\coth{\rm coth}
\def\tanh{\rm tanh}

\def\sqr#1#2{{\vcenter{\vbox{\hrule height.#2pt  
     \hbox{\vrule width.#2pt height#1pt \kern#1pt  
           \vrule width.#2pt}  
       \hrule height.#2pt}}}}  
\def\square{\mathchoice\sqr66\sqr66\sqr{2.1}3\sqr{1.5}3}  
  
\def\appendix{\par\clearpage
  \setcounter{section}{0}
  \setcounter{subsection}{0}
  \def\@sectname{Appendix~}
  \def\theequation{\thesection\arabic{equation}}
  \def\thesection{\Alph{section}}}
 
\def\thefigures#1{\par\clearpage\section*{Figures\@mkboth
  {FIGURES}{FIGURES}}\list
  {Fig.~\arabic{enumi}.}{\labelwidth\parindent\advance
\labelwidth -\labelsep
      \leftmargin\parindent\usecounter{enumi}}}
\def\figitem#1{\item\label{#1}}
\let\endthefigures=\endlist
 
\def\thetables#1{\par\clearpage\section*{Tables\@mkboth
  {TABLES}{TABLES}}\list
  {Table~\Roman{enumi}.}{\labelwidth-\labelsep
      \leftmargin0pt\usecounter{enumi}}}
\def\tableitem#1{\item\label{#1}}
\let\endthetables=\endlist
 
\def\@sect#1#2#3#4#5#6[#7]#8{\ifnum #2>\c@secnumdepth
     \def\@svsec{}\else
     \refstepcounter{#1}\edef\@svsec{\@sectname\csname the#1\endcsname
.\hskip 1em }\fi
     \@tempskipa #5\relax
      \ifdim \@tempskipa>\z@
        \begingroup #6\relax
          \@hangfrom{\hskip #3\relax\@svsec}{\interlinepenalty \@M #8\par}
        \endgroup
       \csname #1mark\endcsname{#7}\addcontentsline
         {toc}{#1}{\ifnum #2>\c@secnumdepth \else
x                      \protect\numberline{\csname the#1\endcsname}\fi
                    #7}\else
        \def\@svse=chd{#6\hskip #3\@svsec #8\csname #1mark\endcsname
                      {#7}\addcontentsline
                           {toc}{#1}{\ifnum #2>\c@secnumdepth \else
                             \protect\numberline{\csname the#1\endcsname}\fi
                       #7}}\fi
     \@xsect{#5}}
 
\def\@sectname{}
%
%
\def\eg{\hbox{\it e.g.}}        \def\cf{\hbox{\it cf.}}
\def\etal{\hbox{\it et al.}}
\def\dash{\hbox{---}}
\def\bR{\mathop{\bf R}}
\def\bC{\mathop{\bf C}}
\def\eq#1{{eq. \ref{#1}}}
\def\eqs#1#2{{eqs. \ref{#1}--\ref{#2}}}
\def\lie{\hbox{\it \$}} 
\def\partder#1#2{{\partial #1\over\partial #2}}
\def\secder#1#2#3{{\partial^2 #1\over\partial #2 \partial #3}}
\def\abs#1{\left| #1\right|}
\def\ltap{\ \raisebox{-.4ex}{\rlap{$\sim$}} \raisebox{.4ex}{$<$}\ }
\def\gtap{\ \raisebox{-.4ex}{\rlap{$\sim$}} \raisebox{.4ex}{$>$}\ }
\def\contract{\makebox[1.2em][c]{
        \mbox{\rule{.6em}{.01truein}\rule{.01truein}{.6em}}}}
%
\def\com#1#2{
        \left[#1, #2\right]}
%
%
\def\bentarrow{\:\raisebox{1.3ex}{\rlap{$\vert$}}\!\rightarrow}
\def\longbent{\:\raisebox{3.5ex}{\rlap{$\vert$}}\raisebox{1.3ex}%
        {\rlap{$\vert$}}\!\rightarrow}
\def\onedk#1#2{
        \begin{equation}
        \begin{array}{l}
         #1 \\
         \bentarrow #2
        \end{array}
        \end{equation}
                }
\def\dk#1#2#3{
        \begin{equation}
        \begin{array}{r c l}
        #1 & \rightarrow & #2 \\
         & & \bentarrow #3
        \end{array}
        \end{equation}
                }
\def\dkp#1#2#3#4{
        \begin{equation}
        \begin{array}{r c l}
        #1 & \rightarrow & #2#3 \\
         & & \phantom{\; #2}\bentarrow #4
        \end{array}
        \end{equation}
                }
\def\bothdk#1#2#3#4#5{
        \begin{equation}
        \begin{array}{r c l}
        #1 & \rightarrow & #2#3 \\
         & & \:\raisebox{1.3ex}{\rlap{$\vert$}}\raisebox{-0.5ex}{$\vert$}%
        \phantom{#2}\!\bentarrow #4 \\
         & & \bentarrow #5
        \end{array}
        \end{equation}
                }
\newcommand{\nc}{\newcommand}
\nc{\spa}[3]{\left\langle#1\,#3\right\rangle}
\nc{\spb}[3]{\left[#1\,#3\right]}
\nc{\ksl}{\not{\hbox{\kern-2.3pt $k$}}}
\nc{\hf}{\textstyle{1\over2}}
\nc{\pol}{\varepsilon}
\nc{\tq}{{\tilde q}}
\nc{\esl}{\not{\hbox{\kern-2.3pt $\pol$}}}
\renewcommand{\theequation}{\arabic{section}.\arabic{equation}}
\renewcommand{\arraystretch}{2.5}
\def\R{1\!\!{\rm R}}
\def\Eins{\mathord{1\hskip -1.5pt
\vrule width .5pt height 7.75pt depth -.2pt \hskip -1.2pt
\vrule width 2.5pt height .3pt depth -.05pt \hskip 1.5pt}}
\newcommand{\symb}{\mbox{symb}}
\renewcommand{\arraystretch}{2.5}
\def\GBd12{{\dot G}_{B12}}
\def\mneg{\!\!\!\!\!\!\!\!\!\!}
\def\Mneg{\!\!\!\!\!\!\!\!\!\!\!\!\!\!\!\!\!\!\!\!}
\def\non{\nonumber}
\def\beqn*{\begin{eqnarray*}}
\def\eqn*{\end{eqnarray*}}
\def\sy{\scriptscriptstyle}
\def\footstrut{\baselineskip 12pt}
\def\square{\kern1pt\vbox{\hrule height 1.2pt\hbox{\vrule width 1.2pt
   \hskip 3pt\vbox{\vskip 6pt}\hskip 3pt\vrule width 0.6pt}
   \hrule height 0.6pt}\kern1pt}
\def\np{n_{+}}
\def\nm{n_{-}}
\def\Np{N_{+}}
\def\Nm{N_{-}}
\def\exmn{\Bigl(\mu \leftrightarrow \nu \Bigr)}
\def\slash#1{#1\!\!\!\raise.15ex\hbox {/}}
\def\dint#1{\int\!\!\!\!\!\int\limits_{\!\!#1}}
\def\bra#1{\langle #1 |}
\def\ket#1{| #1 \rangle}
\def\vev#1{\langle #1 \rangle}
\def\rightvac{\mid 0\rangle}
\def\leftvac{\langle 0\mid}
\def\dps{\displaystyle}
\def\sy{\scriptscriptstyle}
\def\half{{1\over 2}}
\def\third{{1\over3}}
\def\fourth{{1\over4}}
\def\fifth{{1\over5}}
\def\sixth{{1\over6}}
\def\seventh{{1\over7}}
\def\eigth{{1\over8}}
\def\ninth{{1\over9}}
\def\tenth{{1\over10}}
\def\pa{\partial}
\def\ddtau{{d\over d\tau}}
\def\ie{\hbox{$\textstyle{\int_1}$}}
\def\iz{\hbox{$\textstyle{\int_2}$}}
\def\id{\hbox{$\textstyle{\int_3}$}}
\def\ldop{\hbox{$\lbrace\mskip -4.5mu\mid$}}
\def\rdop{\hbox{$\mid\mskip -4.3mu\rbrace$}}
\def\eps{\epsilon}
\def\epshalf{{\epsilon\over 2}}
\def\e{\mbox{e}}
\def\mn{{\mu\nu}}
\def\exmn{{(\mu\leftrightarrow\nu )}}
\def\ab{{\alpha\beta}}
\def\exab{{(\alpha\leftrightarrow\beta )}}
\def\g{\mbox{g}}
\def\kinb{{1\over 4}\dot x^2}
\def\kinf{{1\over 2}\psi\dot\psi}
\def\expk{{\rm exp}\biggl[\,\sum_{i<j=1}^4 G_{Bij}k_i\cdot k_j\biggr]}
\def\expp{{\rm exp}\biggl[\,\sum_{i<j=1}^4 G_{Bij}p_i\cdot p_j\biggr]}
\def\expshort{{\e}^{\half G_{Bij}k_i\cdot k_j}}
\def\expabb{{\e}^{(\cdot )}}
\def\epseps#1#2{\varepsilon_{#1}\cdot \varepsilon_{#2}}
\def\epsk#1#2{\varepsilon_{#1}\cdot k_{#2}}
\def\kk#1#2{k_{#1}\cdot k_{#2}}
\def\G#1#2{G_{B#1#2}}
\def\Gp#1#2{{\dot G_{B#1#2}}}
\def\GF#1#2{G_{F#1#2}}
\def\Dab{{(x_a-x_b)}}
\def\Dsq{{({(x_a-x_b)}^2)}}
\def\lag{( -\partial^2 + V)}
\def\PITD{{(4\pi T)}^{-{D\over 2}}}
\def\4piTD{{(4\pi T)}^{-{D\over 2}}}
\def\4piT4{{(4\pi T)}^{-2}}
\def\TintmD{{\dps\int_{0}^{\infty}}{dT\over T}\,e^{-m^2T}
    {(4\pi T)}^{-{D\over 2}}}
\def\Tintm4{{\dps\int_{0}^{\infty}}{dT\over T}\,e^{-m^2T}
    {(4\pi T)}^{-2}}
\def\Tintm{{\dps\int_{0}^{\infty}}{dT\over T}\,e^{-m^2T}}
\def\Tint{{\dps\int_{0}^{\infty}}{dT\over T}}
\def\pint{{\dps\int}{dp_i\over {(2\pi)}^d}}
\def\Dx{\dps\int{\cal D}x}
\def\Dy{\dps\int{\cal D}y}
\def\Dpsi{\dps\int{\cal D}\psi}
\def\Tr{{\rm Tr}\,}
\def\tr{{\rm tr}\,}
\def\sumij{\sum_{i<j}}
\def\freeexp{{\rm e}^{-\int_0^Td\tau {1\over 4}\dot x^2}}
\def\arraystretch{2.5}
\def\Ge{\mbox{GeV}}
\def\dA{\partial^2}
\def\DA{\sqsubset\!\!\!\!\sqsupset}
\def\FFdual{F\cdot\tilde F}
\def\mn{{\mu\nu}}
\def\rs{{\rho\sigma}}
\def\oplusotimes{{{\lower 15pt\hbox{$\scriptscriptstyle \oplus$}}\atop{\otimes}}}
\def\perppar{{{\lower 15pt\hbox{$\scriptscriptstyle \perp$}}\atop{\parallel}}}
\def\oopp{{{\lower 15pt\hbox{$\scriptscriptstyle \oplus$}}\atop{\otimes}}\!{{\lower 15pt\hbox{$\scriptscriptstyle \perp$}}\atop{\parallel}}}
%
%
\def\bbbr{{\rm I\!R}}
\def\bbbone{{\mathchoice {\rm 1\mskip-4mu l} {\rm 1\mskip-4mu l}
{\rm 1\mskip-4.5mu l} {\rm 1\mskip-5mu l}}}
\def\bbbz{{\mathchoice {\hbox{$\sf\textstyle Z\kern-0.4em Z$}}
{\hbox{$\sf\textstyle Z\kern-0.4em Z$}}
{\hbox{$\sf\scriptstyle Z\kern-0.3em Z$}}
{\hbox{$\sf\scriptscriptstyle Z\kern-0.2em Z$}}}}

\renewcommand{\thefootnote}{\protect\arabic{footnote}}
%

\begin{center}
{\huge\bf }
\vspace{5pt}

{\huge\bf Integral representations combining ladders and crossed-ladders}
\vskip1.3cm

{\large F. Bastianelli$^{a}$, A. Huet$^{b,c}$, C. Schubert$^{b}$, R. Thakur$^{b}$, A. Weber$^{b}$}
\\[1.5ex]

\begin{itemize}

\item [$^a$]
{\it
Dipartimento di Fisica ed Astronomia, Universit\`a di Bologna \\
{\rm and} \\
INFN, Sezione di Bologna, Via Irnerio 46, I-40126 Bologna, Italy.
}

\item [$^b$]
{\it 
Instituto de F\'{\i}sica y Matem\'aticas,
\\
Universidad Michoacana de San Nicol\'as de Hidalgo,\\
Edificio C-3, Apdo. Postal 2-82,\\
C.P. 58040, Morelia, Michoac\'an, M\'exico\\
}

\item [$^c$]
{\it
Departamento de Nanotecnolog\1a,
Centro de F\1sica Aplicada y Tecnolog\1a Avanzada,
Universidad Nacional Aut\'onoma de M\'exico,\\ 
Campus Juriquilla, Boulevard Juriquilla 3001,\\
C.P. 76230, A.P. 1-1010, Juriquilla,Ê Qro., M\'exico.

}

\end{itemize}
\end{center}
\vspace{1cm}
 {\large \bf Abstract:}
\begin{quotation}
We use the worldline formalism to derive integral representations for three classes of amplitudes in scalar
field theory: (i) the scalar propagator exchanging $N$ momenta with a scalar background field (ii) the ``half-ladder'' with $N$ rungs
in $x$ - space
(iii) the four-point ladder with $N$ rungs in $x$ - space as well as in (off-shell) momentum space. 
In each case we give a compact expression combining the $N!$ Feynman diagrams contributing to the amplitude. 
As our main application, we reconsider the well-known case of two massive scalars interacting through
the exchange of a massless scalar.  Applying asymptotic estimates and a saddle-point approximation to the $N$-rung ladder plus crossed ladder
diagrams, we derive a semi-analytic approximation formula for the lowest bound state mass in this model.

 \end{quotation}
\vfill\eject
\pagestyle{plain}
\setcounter{page}{1}
\setcounter{footnote}{0}

\vspace{10pt}
\section{Introduction}
\label{sec:intro}
\renewcommand{\theequation}{1.\arabic{equation}}
\setcounter{equation}{0}

At about the same time when Feynman developed the modern  approach to 
perturbative QED, based on Feynman diagrams, he also invented an alternative representation of the QED effective action or S-matrix in terms of
first-quantized relativistic particle path integrals \cite{feynman:pr80,feynman:pr84}. For the simplest case, the one-loop effective action induced in
scalar QED by an external Maxwell field $A$, this representation reads

\bear
\Gamma [A] &=&
\int_0^{\infty}{dT\over T}\,{\rm e}^{-m^2T}
{\displaystyle \int_{x(T)=x(0)}}{\cal D}x(\tau)
\, e^{-\int_0^T d\tau 
[ \fourth \dot x^2 + ie \dot x^{\mu}A_{\mu}(x(\tau)) ]}
\label{Gammascal}
\ear
Here $T$ denotes the proper-time of the scalar particle in the loop,
$m$ its mass, and $ \int_{x(T)=x(0)}{\cal D}x(\tau)$ a path integral over
all closed loops in spacetime with fixed periodicity in the proper-time
(we will use euclidean conventions throughout this paper). Photon amplitudes as usual
are obtained by specializing the effective action 
to backgrounds involving a finite number of plane waves.

This formalism, which nowadays goes under various
names, e.g. ``Feynman-Schwinger representation'', ``particle presentation'',
``quantum mechanical path integral formalism'', ``first-quantized formalism'' or ``worldline formalism'' 
(which we will adopt here) has been studied by many authors, and extended to other field theories 
(see \cite{41} for an extensive bibliography), but for several decades was  considered as mainly of conceptual interest. 
However, partly as a consequence of developments in string theory \cite{berkos}, where first-quantized
methods figure more prominently than in ordinary field theory, it has in recent years emerged also as a powerful practical tool for the
computation of a wide variety of quantities in quantum field theory. This includes one-loop on-shell \cite{strassler,Bastianelli:1992ct,mckeon:ap224,5} 
and off-shell \cite{strassler2,92} photon/gluon amplitudes,  one- and two-loop 
Euler-Heisenberg-Weisskopf Lagrangians \cite{5,18}, heat-kernel coefficients \cite{6,gussho1}, 
Schwinger pair creation in constant \cite{afalma} and non-constant fields \cite{giekli,63}, Casimir energies \cite{gielan}, 
various types of anomalies (see \cite{basvan-book} and refs. therein), QED/QCD bound states \cite{bravai,nietjoPRL,sagrtj},
heavy-quark condensates \cite{antonov}, and QED/QCD instantaneous Hamiltonians \cite{simonov2013}.  
Extensions to curved space \cite{Bastianelli:2002fv} and quantum gravity \cite{Bastianelli:2013tsa} have also been considered.

\begin{figure}[h]
\centering
\includegraphics[scale=1.4,height=0.1\textheight]{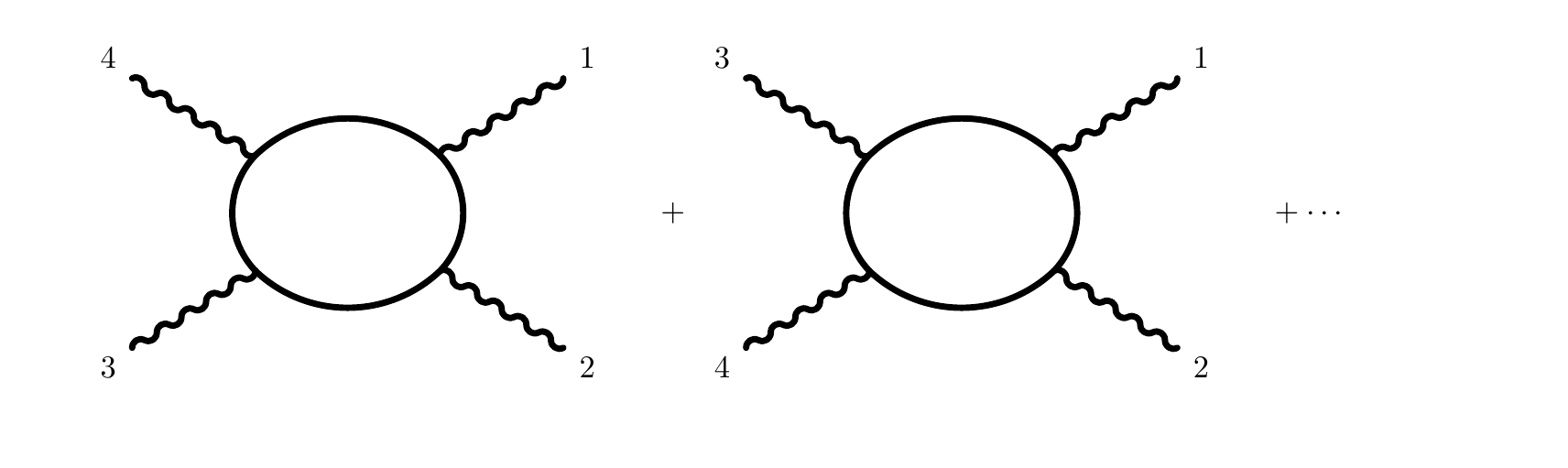}
 \caption{Six permuted diagrams contributing to QED photon-photon scattering.}
 \label{fig:photonphoton}
\end{figure}

One of the interesting aspects of this approach is that often it combines into a single expression 
contributions from a large number of Feynman diagrams. For example, in the QED case it generally allows
one to combine into one integral all contributions from Feynman diagrams which can be
identified by letting photon legs slide along scalar/fermion loops or lines. Thus e.g. the well-known sum of
six permuted diagrams for one-loop QED photon-photon scattering (see fig. \ref{fig:photonphoton}) here naturally
appears combined into a single integral \cite{41}. 

\begin{figure}[h]
\includegraphics[height=.1\textheight]{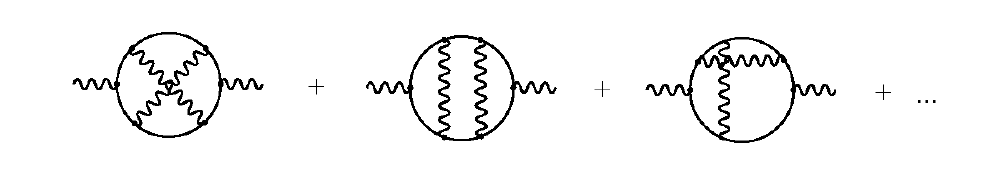}
 \caption{Diagrams contributing to the three loop QED photon propagator.}
 \label{fig-3loopphotonprop}
\end{figure}

While in this case the summation involves graphs that differ only by permutations
of the external legs, at higher loop orders the summation will generally involve
topologically different diagrams;  as an example,
we show in fig. \ref{fig-3loopphotonprop} the ``quenched'' contributions to the 
three-loop photon propagator.

This property is particularly interesting in view of the fact that it is just
this type of summation which in QED often leads to extensive cancellations,
and to final results which are substantially simpler than intermediate 
ones (see, e.g., \cite{cvitanovic1977,brdekr}).  
More recently, similar cancellations have been found also for graviton amplitudes (see, e.g., \cite{babiva}).

\begin{figure}[h]
\centering
\includegraphics[scale=.8,height=.1 \textheight]{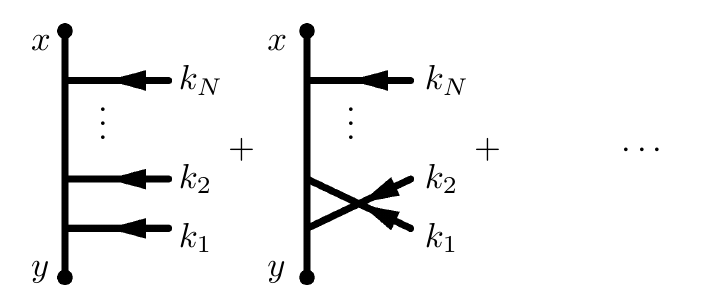}
\caption{Sum of diagrams contributing to the $N$ - propagator.}
\label{fig:Nprop}
\end{figure}

Although this property of the worldline formalism is well-known, and has been occasionally exploited \cite{rossch,15,barroberg:fbs} 
(see also \cite{frigab-summation})
a systematic study of its implications is presently still lacking.
In this paper, we will initiate such a study for the simplest case of scalar field theory, considering two real scalar
fields interacting through a cubic vertex. In this model, we will look at the following
three classes of Green's functions: the first one, depicted in fig. \ref{fig:Nprop}, is the $x$-space
propagator for one scalar interacting with the second one through the exchange of $N$ given momenta.

This object, to be called ``$N$-propagator'', is given by a set of $N!$ simple tree-level graphs,
and in section \ref{sec:propagators} we will use the worldline formalism to combine them into
a single integral. We will also obtain the momentum-space version of this result. 

\begin{figure}[ht]
\centering
\includegraphics[scale=.8]{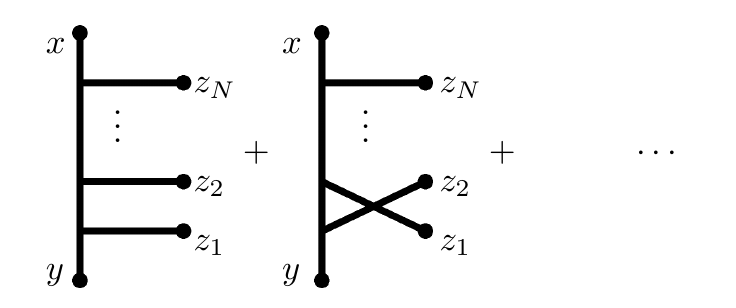}
\caption{Sum of diagrams contributing to the half-ladder.}
\label{fig:halfladder}
\end{figure}

The second class are the similarly looking  $x$-space $N+2$ - point functions shown in fig \ref{fig:halfladder},
defined by a line connecting the points $x$ and $y$ and $N$ further points $z_1,\ldots,z_N$ connecting to this
line in an arbitrary order. 

These ``$N$-rung half-ladders'' again form a set of $N!$ diagrams, and we will give a unifying integral
representation in section \ref{sec:half-ladder}. This class of diagrams is, apart from the first ($N=1$)
one, which is just the well-known off-shell scalar triangle integral \cite{ussdav}, already highly nontrivial;
the four-point integral corresponding to $N=2$ figures prominently in $N=4$ SYM theory \cite{36,bkpss,henhub,bogrpr} 
(it was called $f(x_1,x_2,x_3,x_4)$ in \cite{36}) but is presently
still not known in closed form. Here we will derive for it a novel two-parameter integral representation.

\begin{figure}[ht]
\centering
\includegraphics[scale=.8]{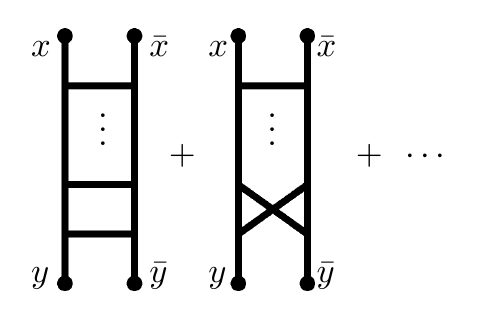}
\caption{Sum of ladder and crossed-ladder contributions to the four-point function in $x$ - space.}
\label{fig:ladders}
\end{figure}

Finally, in section \ref{sec:ladders}
we come to the class of ladder graphs, depicted in fig.\ \ref{fig:ladders}, which
we obtain by ``gluing together'' two ``$N$-propagators''.

Just as in the case of the $N$-propagators and 
half-ladders, one distinctive advantage of the worldline representation
over the usual Feynman parameterization of this type of diagrams is the
automatic inclusion of all possible ways of crossing the ``rungs'' of the
ladders. Here again we will obtain such unifying representations in explicit
form both in $x$-space and in momentum space.

Ladder graphs with a finite number $N$ of rungs play an important role for
scattering processes in the high energy, large momentum transfer limit, see,
\eg, ref.\ \cite{gastro:plb249}. In this paper, we will concentrate on the
case of infinite $N$, \hbox{\it i.e.}, the sum over {\it all} ladder {\it and}
crossed ladder graphs, which is of paramount importance for the bound state
problem. In fact, our hope that a fresh look at these graphs from the
perspective of the worldline formalism, usually refered to as the worldline
representation in this context (see, e.g.,\cite{simtjo:ap228}),
can give new insights in the bound state problem is the original motivation 
behind the present work.

It is our opinion that the bound state problem, in the sense of establishing
an efficient and systematic formalism that would allow one to calculate the
bound states and their properties for a given field theory, is one of the
important open problems in quantum field theory, and that the fact that
so little work is dedicated at present to this problem reflects its
complexity rather than a lack of importance. It is evident, in fact,
that the present-day description of (light) hadrons, which are intrinsically
relativistic bound states of quarks and gluons, is not satisfactory from a 
theoretical standpoint. Not only a precise description of the effective
interaction of quarks and gluons is missing, but also a convenient formalism
for the calculation of the hadronic states once an appropriate description
of the interaction is established.

This being said, a fully relativistic equation for the masses and structure
of the bound states of two constituents
has been established in quantum field theory a long time ago 
by Salpeter and Bethe \cite{salbet:pr84,gellow:pr84}. Unfortunately, the
practical application of this equation suffers from all kinds of difficulties,
see, \eg, ref.\ \cite{nakanishi:ptps43} for an early review. In particular,
despite the fact that the equation is exact in principle, applications can 
hardly go beyond the ladder approximation to the equation which amounts to
replacing the totality of diagrams contributing to the four-point function
with the ladder graphs, {\it excluding} all crossed ladder graphs. The
inclusion of the crossed ladder graphs, however, is essential for the
consistency of the one-body limit where one of the constituents becomes
infinitely heavy, and for maintaining gauge invariance (in gauge theories).

Alternatives to the Bethe-Salpeter equation have been devised that partially 
include the crossed ladder graphs, the best-known being the
Blan\-ken\-bec\-ler-Sugar equation \cite{blasug:pr142,logtav:nc29}, the Gross 
(or spectator) equation \cite{gross:pr186} and the equal-time equation 
\cite{walman:npa503}. In order to assess how well these so-called
quasipotential equations are doing in incorporating the effects of the
crossed-ladder graphs, and to establish some benchmark values for the
relativistic bound state problem, Nieuwenhuis and Tjon \cite{nietjoPRL} have numerically
evaluated the path integrals of the worldline representation for
the same scalar model field theory that we are considering here, thus including all ladder
{\it and} crossed ladder graphs. The results, if the numerical
evaluation is to be trusted, are not reassuring: while the predictions of
the quasi-potential equations are closer to the numerical values for the
lowest bound state mass than the solution of the Bethe-Salpeter equation, they
still differ substantially from the worldline values (and from one
another). On the other hand, the predictions of the quasipotential equations 
for the equal-time wave function of the lowest bound state are {\it worse}
than the ones of the Bethe-Salpeter equation.
Similar conclusions concerning the importance of crossed contributions
were reached for the same model in the more extensive study by Savkli et al. \cite{sagrtj}. Here
both numerical and analytical methods were used in the evaluation of the
worldline path integrals, and some results were obtained also for 1+1 dimensional Scalar QED.

In section \ref{sec:wc}, we will apply the worldline representation
to the same scalar model field theory that was considered by Nieuwenhuis and 
Tjon, but we will derive concrete results for the mass of the lowest bound state
for the case of a massless exchanged particle (along the ``rungs'' of the
ladders), while Nieuwenhuis and Tjon took the mass of the exchanged particle
to be $0{.}15$ times the mass of the constituents. Furthermore, we are
interested in exploring how far one can get in an (approximate) analytical,
rather than numerical, evaluation of the path integrals.

We should also like to mention that, particularly in the case of a massless
exchanged particle, field theoretical perturbation theory can be applied
in order to calculate corrections to the essentially nonrelativistic
situation, as long as the coupling constant is sufficiently small. In this
way, very precise predictions have been obtained for the case of positronium.
For comparison, if one applies the Bethe-Salpeter equation in the ladder
approximation to the scalar model field theory with a massless exchanged 
particle, known in this context as the Wick-Cutkosky model \cite{wick,cutkosky}, the bound state
solutions tend to their nonrelativistic counterparts (the interaction of the
constituents being described by a Coulomb potential) in the nonrelativistic
limit of small coupling constant. However, already the first relativistic
corrections (in an expansion in powers of the coupling constant) as predicted 
by the Bethe-Salpeter ladder approximation are considered unphysical
\cite{felful:prd7,itzzub-book}. We will return to this issue in section \ref{sec:wc}.

Now let us define our model. We will work in the euclidean throughout in this paper.
The action for our field theory with two
scalars interacting through a cubic vertex is

\bea
S[\phi,\chi] = \int d^Dx \, \bigg ( {1\over 2} ( \partial_\mu \phi)^2  + {1\over 2}  m^2\phi^2  
+ {1\over 2} ( \partial_\mu \chi)^2  
+ {1\over 2}  \mu^2\chi^2
+{\lambda\over 2!} \phi^2\chi \bigg ) \ .
\nonumber\\
\label{S}
\eea
Our most basic object of interest is the propagator for the $\phi$ - field in the
background of the $\chi$ - field. The worldline representation of this 
propagator is (for a careful derivation see \cite{holten:npb457})

\bea
\la 0|T \phi (x) \phi (y) |0\ra_{(\chi)} 
= \int_0^\infty \! dT \,\e^{-m^2 T} \int_{_{x(0)=y}}^{^{\, x(T)=x}} 
\!\!\!\!\!\!\!\!\!\!\!\! {\cal D}x \
e^{ -\int_0^Td\tau\, \big [{1\over 4} \dot x^2 + \lambda\chi(x(\tau)) \big]} \ .
\label{wp}
\eea
Here the path integral runs over all trajectories in euclidean space that lead
from $y$ to $x$  in the fixed proper-time $T$. 
From this propagator in the background field we can obtain the ``$N$-propagator''
for the $\phi$ - particle, describing its interaction with the $\chi$ - field through the interchange of $N$ 
quanta with four-momenta $k_1,\ldots,k_N$. This simply requires specializing the background scalar field $\chi(x)$ 
to a sum of $N$ plane waves,

\bear
\chi(x) = \sum_{i=1}^N \e^{ik_i\cdot x}
\label{chisum}
\ear
and picking the terms linear in each of the plane waves on the rhs of (\ref{wp}).
For the $N$-propagator (\ref{wp}) induces the representation

\bear
\la 0|T \phi (x) \phi (y) |0\ra_{(N)} &=& (-\lambda)^N
 \int_0^\infty \! dT \,\e^{-m^2 T} 
 \int_0^Td\tau_1 \cdots \int_0^Td\tau_N
 \nonumber\\ &&
\times  \int_{_{x(0)=y}}^{^{\, x(T)=x}} 
\!\!\!\!\!\!\!\!\!\!\!\! {\cal D}x 
\,\e^{i\sum_{i=1}^Nk_i\cdot x(\tau_i)}
\e^{ -\int_0^Td\tau\, {1\over 4} \dot x^2} \ .
\nonumber\\
\label{Nprop}
\eea
The path integral is now of Gaussian type, so that it can be evaluated exactly using
only the determinant and the inverse (``worldline Green function'') of the kinetic operator, which here is simply the second derivative operator in proper-time
. In section \ref{sec:propagators} we will do this in detail. 
For the scalar field theory amplitudes considered in this paper, the resulting ``worldline integrals'' are related to standard Feynman 
parameter integrals in a straightforward way. However, they offer an advantage over Feynman parameter integrals in that
they are valid independently of the ordering of the momenta $k_1,\ldots,k_N$; the rhs of (\ref{Nprop}) contains already all $N!$ 
possibilities of attaching the $N$ momenta to the propagator, as shown in figure \ref{fig:Nprop}.

Although all the integrals considered in this paper are finite in four dimensions, we will work in a general
dimension $D$, except in some of our more explicit calculations. 

\section{$N$-propagators}
\label{sec:propagators}
\renewcommand{\theequation}{2.\arabic{equation}}
\setcounter{equation}{0}

We proceed to the calculation of the Gaussian path integral (\ref{Nprop}). 
First, let us split  $x^\mu(\tau)$ into a background part $x^\mu_{bg}(\tau)$,
which encodes the boundary conditions, and a quantum part  $q^\mu(\tau)$, 
which has zero Dirichlet boundary conditions at $\tau=0,T$,

\bea 
x^\mu(\tau) &=& x^\mu_{bg}(\tau)+ q^\mu(\tau)\, , \nonumber\\
x^\mu_{bg}(\tau) &=& 
y^\mu + (x-y)^\mu \frac{\tau}{T} \, , \nonumber\\
q^\mu(0)=q^{\mu}(T) &=&  0 \, . \nonumber\\
\label{split}
\eea
The propagator for $q^\mu(\tau)$  is the Green's function for the
second derivative operator on an interval of length $T$ with vanishing boundary conditions,
which is \cite{fuoswi,Bastianelli:1992ct,mckeon:ap224}

\bea 
\la q^\mu(\tau)  q^\nu(\sigma) \ra 
&=&
-2 \delta^{\mu\nu} \Delta_T(\tau,\sigma) \, , \nonumber\\
\Delta_T(\tau,\sigma) &=&
\frac{\tau\sigma}{T} -\tau \theta(\sigma-\tau) 
- \sigma\theta(\tau -\sigma)
\nonumber\\
&=& \frac{\tau\sigma}{T} +\frac{\vert\tau - \sigma \vert}{2} - \frac{\tau + \sigma}{2}
\, .
 \nonumber\\ 
\label{defDelta}
\eea
We note also the coincidence limit of this Green's function,

\bear
\Delta_T(\tau,\tau) = \frac{\tau^2}{T} -\tau \ . \nonumber\\
\label{coinDelta}
\eea
We will also need the free path integral normalization factor (see, e.g. \cite{basvan-book})

\bear
 \int_{_{q(0)=0}}^{^{\, q(T)=0}} 
\!\!\!\!\!\!\!\!\!\!\!\! {\cal D}q
\,\e^{ -\int_0^Td\tau\, {1\over 4} \dot q^2} 
=
\frac{1}{(4\pi T)^{D\over 2}}
\, .
\label{freepi}
\ear

\begin{figure}[h]
\centering
\includegraphics{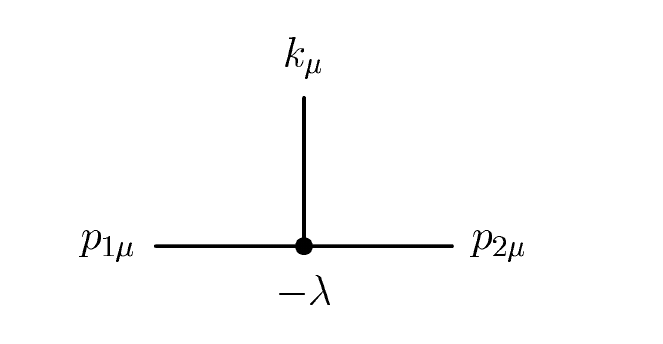}
\caption{Scalar vertex}
\label{fig:vertex}
\end{figure}

For the benefit of the reader unfamiliar with worldline path integrals,
let us first consider the case $N=1$. From (\ref{Nprop}), (\ref{split}) and (\ref{freepi})

\bea
\la 0|T \phi (x) \phi (y) |0\ra_{_{\!(1)}} =
\int_0^\infty \! {dT\over (4\pi T)^{D\over 2}}\, 
e^{  - {(x-y)^2\over 4T} - m^2T  }
(-\lambda)\int_0^Td\tau\, e^{ik \cdot [y+(x-y)\frac{\tau}{T}]}\la e^{ik \cdot q(\tau)}\ra 
\nonumber\\
\label{1propsplit}
\eea
with the Wick contraction

\bea
\la e^{ik \cdot q(\tau)}\ra = e^{k^2 (\frac{\tau^2}{T} -\tau)}
\label{wick1}
\eea 
by (\ref{coinDelta}). Summarizing,

\bea
\la 0|T \phi (x) \phi (y) |0\ra_{_{\!(1)}} =
\int_0^\infty \! {dT\over (4\pi T)^{D\over 2}}\, 
e^{  - {(x-y)^2\over 4T} - m^2T  }
(-\lambda)\int_0^Td\tau\,
\underbrace{e^{ik \cdot [y+(x-y)\frac{\tau}{T}]}}_{classical\ path}
\, 
\underbrace{e^{- k^2 (\tau -\frac{\tau^2}{T})}}_{Wick\ contr.}
\, .
 \nonumber\\
\label{1prop}
\eea
We Fourier transform in $x$ and $y$, rescale $\tau = Tu$, do the $T$ integral
and obtain the product of two propagators in the Feynman parametrization

\bea
\la \tilde \phi(p_1) \tilde \phi(p_2)\ra_{_{\!(1)}} 
\!\!&=&\!\! 
(2\pi)^D \delta^D(p_1+p_2+k) (-\lambda)
\int_0^1 du\,
\int_0^\infty dT\, T\, e^{-T [p_1^2 +m^2 + (k^2+2p_1\cdot k )u ]} 
\nonumber \\
&=&\!\! 
(2\pi)^D \delta^D(p_1+p_2+k)\, (-\lambda) 
\int_0^1 du\, {\Gamma(2)\over [p_1^2 + m^2 + (k^2 +2 p_1\cdot k)u]^2}
\nonumber \\
&=&\!\! 
(2\pi)^D \delta^D(p_1+p_2+k)\,
{1\over p_1^2 + m^2}\,  (-\lambda) \,  {1\over (p_1+k)^2 + m^2} \ .
\label{svertex}
\eea
Thus we have recovered the standard Feynman rule expression for 
the basic scalar vertex (fig. \ref{fig:vertex}).

\no
Proceeding directly to the $N$-point case, here (\ref{1propsplit}) generalizes to 

\bea
\la 0|T \phi (x) \phi (y) |0\ra_{_{\!(N)}} &=&
\int_0^\infty \! {dT\over (4\pi T)^{D\over 2}}\, 
e^{  - {(x-y)^2\over 4T} - m^2T  }
(-\lambda)^N\int_0^Td\tau_1\cdots \int_0^Td\tau_N \nonumber\\
&&\times 
{\rm e}^{i\sum_i k_i\cdot \bigl(y+ \frac{\tau_i}{T}(x-y)\bigr)}
 \la e^{i\sum_{i=1}^N k_i \cdot q(\tau_i)}\ra \, .
\nonumber\\
\label{Npropsplit}
\eea
After performing the Gaussian integration over $q^{\mu}(\tau)$ using the Green's function 
(\ref{defDelta}), and a rescaling $\tau_i = Tu_i$, this becomes

\bea
\la 0|T \phi (x) \phi (y) |0\ra_{_{\!(N)}} 
&=&
(-\lambda)^N
\int_0^\infty \! { dT 
\over (4\pi T)^{D\over 2}}\, 
e^{  - {(x-y)^2\over 4T} - m^2T}
T^N 
\int_0^1 du_1 \cdot\cdot\int_0^1du_N\, 
\nonumber\\
&&\times
{\rm e}^{i\sum_i k_i\cdot \bigl(y+ u_i(x-y)\bigr)}
{\rm exp}\biggl\lbrack
T\sum_{i,j=1}^Nk_i\cdot k_j \Delta_1(u_i,u_j)
\biggr\rbrack
\, .
\nonumber\\
\label{orderNxspace}
\eea
Fourier transformation of this representation yields, after an easy computation,

\bea
\la \tilde \phi(p_1) \tilde \phi(p_2)\ra_{_{\!(N)}} 
&=&
(2\pi)^D \delta^D(p_1+p_2+ \sum_i k_i) (-\lambda)^N 
\int_0^{\infty}dT
\int_0^T d\tau_1 \cdot\cdot\int_0^Td\tau_N\, 
\nonumber\\
&& \times {\rm exp} \biggl\lbrace - T \Bigl( p_1^2 +m^2\Bigr) 
- \sum_i (k_i^2+2p_1\cdot k_i )\tau_i
- \sum_{i<j} 2k_i \cdot k_j D(\tau_i,\tau_j)
\biggr\rbrace \nonumber\\
&=& 
(2\pi)^D \delta^D(p_1+p_2+ \sum_i k_i) (-\lambda)^N  N! 
\int_0^1 du_1 \cdot\cdot\int_0^1du_N\, 
\nonumber\\
&& \hspace{-30pt} \times  \bigg [p_1^2 +m^2 + \sum_i 
(k_i^2+2p_1\cdot k_i )u_i
+ \sum_{i<j} 2k_i \cdot k_j 
D(u_i,u_j)\bigg]^{-N-1} \ .
\nonumber\\
\label{orderNpspace}
\eea
where we have further defined

\bea
D(\tau_i,\tau_j) &:=& \tau_i \theta(\tau_j -\tau_i)+ 
\tau_j \theta(\tau_i -\tau_j)
\, .
\label{defD}
\eea
Each of the $N!$ orderings of the $u_1,..,u_N$
parameters along the worldline region $[0,1]$ identifies a
range of integration.
Each range of integration produces the 
product of the $(N+1)$ propagators where the momentum flows
according to momentum conservation.
This gives the total of $N!$ contributions corresponding to the
various exchanges of the external lines carrying momentum $k^\mu_i$.
The explicit proof is given in the appendix.

Our leitmotif in this paper is to find representations like (\ref{orderNxspace}) and
(\ref{orderNpspace}) that unify the Feynman diagrams corresponding to different orderings.
However, as an aside we wish to mention also that the contribution of any ordered sector to (\ref{orderNxspace})
can be recast in a form that is a finite-dimensional analogue of the initial path-integral
(\ref{Nprop}). First, introducing the inverse of the $N\times N$ matrix 
$-\Delta_{ij} = -\Delta_1(u_i,u_j)$, as well as its determinant $|-\Delta|$,
we can trivially rewrite the final exponential factor in (\ref{orderNxspace}) in terms of 
a Gaussian integral over auxiliary variables $\xi_1, \ldots \xi_N$ as

\bear
{\rm exp}\biggl\lbrack
T\sum_{i,j=1}^Nk_i\cdot k_j \Delta_1(u_i,u_j)
\biggr\rbrack
&=&
\int d^D\xi_1 \cdots \int d^D\xi_N
\biggl((4\pi T)^N |-\Delta|\biggr)^{-\frac{D}{2}}
\nonumber\\
&&\times
\,{\rm exp}\biggl[-\frac{1}{4T} \sum_{i,j=1}^N (-\Delta^{-1})_{ij} \xi_i\cdot \xi_j + i \sum_{i=1}^N k_i\cdot \xi_i \biggr]
\, .
\nonumber\\
\label{rewriteexp}
\ear
It is sufficient to consider the standard ordering $1\ge u_1\ge u_2 \ge \ldots \ge u_N \ge 0$.
For this sector, it is straightforward to show inductively that $|-\Delta|$ and $(-\Delta^{-1})$
are given by 

\bear
|-\Delta|
= (1-u_1)(u_1-u_2)(u_2-u_3)\cdots (u_{N-1}-u_N)u_N 
\label{detDelta}
\ear
and

\bear\small
-\Delta^{-1} \!\! = \!\! \tiny
\left(\begin{array}{ccccc}
\frac{1}{1-u_1}+\frac{1}{u_1-u_2} & - \frac{1}{u_1-u_2}  & 0 & 0 & 0 \\
- \frac{1}{u_1-u_2}  & \frac{1}{u_1-u_2}+\frac{1}{u_2-u_3}  &- \frac{1}{u_2-u_3}  & 0 & 0 \\
\vdots & \vdots & \vdots & \vdots & \vdots \\
0 & 0 & - \frac{1}{u_{N-2}-u_{N-1}} & \frac{1}{u_{N-2}-u_{N-1}}+\frac{1}{u_{N-1}-u_N} & - \frac{1}{u_{N-1}-u_N} \\
0 & 0 & 0 & - \frac{1}{u_{N-1}-u_N} & \frac{1}{u_{N-1}-u_N} + \frac{1}{u_N} 
\end{array}\right)
\nonumber\\
\label{invDelta}
\ear
Thus in the first term in the exponent in (\ref{rewriteexp}) we can rewrite

\bear
\sum_{i,j=1}^N (-\Delta^{-1})_{ij} \xi_i\cdot \xi_j 
= 
\frac{\xi_1^2}{1-u_1} + \sum_{i=1}^{N-1}\frac{(\xi_i-\xi_{i+1})^2}{u_i-u_{i+1}} + \frac{\xi_N^2}{u_N}
\label{rewritefirst}
\ear
Using (\ref{rewritefirst}) in (\ref{orderNxspace}) and performing a linear shift 

\bear
\xi_i \to \xi_i - y - u_i(x-y)
\label{shiftxi}
\ear
we get 

\bea
\la 0|T \phi (x) \phi (y) |0\ra_{_{\!(N)}}^{(12\ldots N)} 
\!\! &=& \!\!
(-\lambda)^N
\int_0^\infty \! { dT 
\over (4\pi T)^{D\over 2}}\, 
e^{ - m^2T}
T^N 
\int_0^1 du_1 \int_0^{u_1} du_2 \cdots \int_0^{u_{N-1}}du_N\, 
\nonumber\\
&& \times
\int d^D\xi_1 \cdots \int d^D\xi_N
\biggl((4\pi T)^N |-\Delta|\biggr)^{-\frac{D}{2}}
\nonumber\\
&&\times
{\rm exp}\Biggl\lbrace
- \frac{1}{4T} 
\biggl\lbrack
\frac{(x-\xi_1)^2}{1-u_1} + \sum_{i=1}^{N-1}\frac{(\xi_i-\xi_{i+1})^2}{u_i-u_{i+1}} + \frac{(\xi_N-y)^2}{u_N}
\biggr\rbrack
+ i \sum_{i=1}^N k_i \cdot \xi_i
\Biggr\rbrace
.
\nonumber\\
\label{xirep}
\eea
Here on the lhs the superscript $(12\ldots N)$ indicates the restriction to the standard ordering.
Comparing with the original path integral (\ref{Nprop}) it will be observed that 
(\ref{xirep}) can be viewed as a restriction of this path integral to the finite-dimensional
set of  {\it polygonal} paths leading
from $x$ to $y$, corresponding to the propagation of a particle that is free in between 
absorbing (or emitting), at proper-time $\tau_i = u_iT$ and the space-time point $\xi_i$, the momentum $k_i$.
Alternatively, 
the representation (\ref{xirep}) of the $N$ - propagator can also be obtained
using heat-kernel methods similar to the ones of \cite{fuoswi}. Despite of its simplicity we
have not been able to find this formula in the literature.
%
%

\section{$N$-point half-ladders}
\label{sec:half-ladder}
\renewcommand{\theequation}{3.\arabic{equation}}
\setcounter{equation}{0}

We proceed to the set of half-ladder diagrams depicted in fig. \ref{fig:halfladder}. Those we
will consider in $x$ - space only. They can be obtained from the $N$-propagators by replacing 

\begin{equation}
e^{ik_i\cdot x(\tau_i)} \to 
\int \frac{d^Dk_i}{(2\pi)^D} \frac{e^{ik_i\cdot(x(\tau_i) - z_i)}}{k_i^2 + \mu^2}
\label{proptihl}
\end{equation}
for $i=1,\ldots,N$. 
For $N=1$ we obtain, after this replacement, the usual Schwinger exponentiation

\begin{equation}
\frac{1}{k^2+\mu^2} = \int_0^{\infty}d\alpha\,{\rm e}^{-\alpha(k^2+\mu^2)}
\label{schwinger2}
\end{equation}
and the use of (\ref{1prop}), 
the following representation for the lowest-order scalar $x$-space three-point function, with two propagators
having mass $m$ and one having mass $\mu$:

\begin{eqnarray}
\Gamma(x,y,z,m,\mu) &=& -\lambda
\int_0^\infty \! {dT\over (4\pi T)^{D\over 2}}\, 
e^{  - {(x-y)^2\over 4T} - m^2T  }\int \frac{d^Dk}{(2\pi)^D}\int_0^{\infty}d\alpha\,{\rm e}^{-(ik\cdot z +\alpha(k^2+\mu^2))}
\nonumber\\
&& \times \int_0^Td\tau\,
e^{ik\cdot y} e^{ik \cdot (x-y)\frac{\tau}{T}}
\, 
e^{- k^2 (\tau -\frac{\tau^2}{T})}
\, .
\label{halfladder1}
\end{eqnarray}
Performing the Gaussian $k$-integral and rescaling $\tau = Tu$ as well as $\alpha = T \hat\alpha$, we obtain

\begin{eqnarray}
\Gamma(x,y,z,m,\mu) &=&
-\frac{\lambda}{(4\pi)^D}\int_0^\infty \! \frac{dT} {T^{D - 2}}\, 
e^{  - {(x-y)^2\over 4T} - m^2T}
\int_0^\infty d\hat\alpha \, e^{-\hat\alpha \mu^2T}
\nonumber\\
&& \times
\int_0^{1}du \frac{1}{[\hat\alpha + u(1-u)]^{\frac{D}{2}}} e^{\frac{-(y - z + (x - y)u)^2}{4T(\hat\alpha + u(1-u))}}
\, .
\label{kdone1}
\end{eqnarray}
Now we specialize to the massless case, $m=\mu=0$. The $T$ - integral then becomes elementary, and 
one gets 

\begin{eqnarray}
\Gamma(x,y,z,0,0) &=&
-\frac{\lambda}{(4\pi)^D}\Gamma(D-3)
\int_0^1du\int_0^\infty d\hat\alpha \nonumber\\
&& \times
 \frac{1}{[\hat\alpha + u(1-u)]^{\frac{D}{2}}}
 \frac{4^{D-3}}
  {\Bigl[(x-y)^2+\frac{[y - z + (x - y)u]^2}{\hat\alpha + u(1-u)}\Bigr]^{D-3}}
  \, .
  \nonumber
\end{eqnarray}
Further simplification is possible if we now also assume $D=4$. This makes the
$\hat\alpha$ - integral elementary, and results in

\begin{eqnarray}
\Gamma(x,y,z,0,0) 
&=& 
-\frac{\lambda}{64\pi^4}
\int_0^{1} \!du \frac{1}{uc+(1-u)b-u(1 - u)a} \log\Bigl[\frac{uc+(1-u)b}{u(1 - u)a}\Bigr]
\nonumber\\
\label{nm}
\end{eqnarray}
where we have now abbreviated

\begin{eqnarray}
(x -y)^2 = a\, , \hspace{20pt} (y - z)^2 = b\, , \hspace{20pt} (x - z)^2 = c \, .
\label{defabc}
\end{eqnarray}
The $u$ - integral can be reduced to the standard integral

\bear
\int du \frac{\ln (Au+B)}{u-C} 
&=& \ln (Au+B)\ln \Bigl(1-\frac{Au+B}{AC+B}\Bigr) 
+ {\rm Li}_2 \Bigl(\frac{Au+B}{AC+B}\Bigr) 
\, .
\nonumber\\
\label{standardint}
\ear
The final result is then easy to identify with the well-known representation of the massless triangle 
function due to Ussyukina and Davydychev \cite{ussdav},

\begin{equation}
\Gamma(x,y,z,0,0)=  -\frac{\lambda}{64\pi^4} \frac{1}{a}\Phi^{(1)}\Bigl(\frac{b}{a}, \frac{c}{a}\Bigr)
\end{equation}
where

\begin{eqnarray}
\Phi^{(1)}(x,y)
&:= &
{1\over \Lambda}
\Biggl\lbrace
2\Bigl({\rm Li}_2(-\rho x)
+
{\rm Li}_2(-\rho y)\Bigr)
+\ln
{y\over x}
\ln
{{1+\rho y}\over {1+\rho x}}
+
\ln (\rho x)\ln (\rho y)
+
{\pi^2\over 3}
\Biggr\rbrace
\nonumber\\
\label{defPhi1}
\end{eqnarray}
with

\bear
\Lambda &:=& \sqrt{(1-x-y)^2-4xy},
\nonumber\\
\rho &:=& 2(1-x-y+\Lambda)^{-1}.
\nonumber\\
\label{defLambdarho}
\ear
After this warm-up, we proceed to the much more challenging $N=2$ case. Eq. (\ref{halfladder1}) generalizes 
straightforwardly to

\begin{eqnarray}
\Gamma (x,y,z_1,z_2,m,\mu) &=& (-\lambda)^2\int^\infty_0 \frac{dT\, T^2}{(4\pi T)^{D/2}} e^{-\frac{(x - y)^2}{4T} -m^{2} T} 
\int \frac{d^Dk_1}{(2\pi)^D}\int \frac{d^Dk_2}{(2\pi)^D} \,e^{-i(k_{1}\cdot z_{1} + k_{2}\cdot z_{2})}
\nonumber
\\
&&
\times
\int _0^{\infty} d\alpha_{1}e^{-\alpha_1(k_1^2 + \mu^2)} \int _0^{\infty} d\alpha_2 \,e^{-\alpha_{2}(k_{2}^2 + \mu^2)}
\int^{1}_{0} du_{1}\int^{1}_{0} du_{2} 
\nonumber\\
&&\times
e^{ik_{1}\cdot[y + (x - y)u_1]}
e^{ik_{2}\cdot[y + (x - y)u_2]} 
e^{T\bigr[k_{1}^2\bigtriangleup_{1}(u_{1},u_{1}) + k_{2}^2\bigtriangleup_{1}(u_2,u_2)  +2k_{1}\cdot k_{2}\bigtriangleup_{1}(u_1,u_2)\bigr]}
\, .
\nonumber\\
\label{starting2}
\end{eqnarray}
Here we have already rescaled $\tau_i = Tu_i$, $i=1,2$. The ordered sector $u_1 < u_2$ of this integral corresponds to the first diagram shown in
fig. \ref{fig:halfladder} (for $N=2$), the sector $u_1>u_2$ to the second one. 

As before, we first do the
Gaussian $k_{1,2}$ - integrals, and obtain 
(in the following we abbreviate $\bigtriangleup_1 (u_i,u_j)$ by $\bigtriangleup_{ij}$)

\begin{eqnarray}
\Gamma (x,y,z_1,z_2,m,\mu) &=& \frac{\lambda^2}{(4\pi)^D}\int^\infty_0 \frac{dT\, T^2}{(4\pi T)^{D/2}} e^{-\frac{(x - y)^2}{4T} -m^{2} T} 
\int^{1}_{0} du_{1}du_2
\int_0^{\infty} d\alpha_{1} d\alpha_{2}
e^{-(\alpha_1 + \alpha_{2}) \mu^2}
\nonumber
\\
&&\times
\frac{{\rm exp}\Bigl\lbrace -\frac{(\alpha_1-T\bigtriangleup_{11}) \beta_2^2 + (\alpha_2-T\bigtriangleup_{22}) \beta_1^2 + 2T \bigtriangleup_{12}\beta_1\cdot \beta_2}
{4 \bigl[(\alpha_1-T\bigtriangleup_{11})(\alpha_2-T\bigtriangleup_{22}) - T^2 \bigtriangleup_{12}^2\bigr]}
\Bigr\rbrace
}{\Bigl[(\alpha_1-T\bigtriangleup_{11})(\alpha_2-T\bigtriangleup_{22}) - T^2 \bigtriangleup_{12}^2\Bigr]^{\frac{D}{2}}}
\, 
\nonumber\\
\label{Gammafourkdone}
\end{eqnarray}
where we have defined

\bear
\beta_i := y - z_i + u_i (x-y) \, .
\label{defbetai}
\ear
Specializing to the massless case $m=\mu=0$, and changing from $\alpha_i$ to $\hat\alpha_i$ via

\bear
\alpha_i = T (\hat\alpha_i + \bigtriangleup_{ii}),\quad  i=1,2,
\label{defalphahati}
\ear
we can do the $T$ - integral. This leads to

\begin{eqnarray}
\Gamma (x,y,z_{1},z_{2},0,0) &=& \frac{\lambda^2}{(4\pi)^{\frac{3}{2}D}}\Gamma\Bigl(1+\frac{3}{2}(D-4)\Bigr) 
\int^{1}_{0} du_{1}du_2
\int_{-\bigtriangleup_{11}}^{\infty} d\hat\alpha_{1} 
\int_{-\bigtriangleup_{22}}^{\infty} d\hat\alpha_{2} 
\nonumber\\ 
&&\times 
\frac{1}{\Bigl[\hat\alpha_1 \hat\alpha_2 - \Delta_{12}^2\Bigr]^{\frac{D}{2}}}
\Biggl\lbrack
\frac{4}
{(x-y)^2 + \frac{\hat\alpha_1\beta_2^2 + \hat\alpha_2 \beta_1^2 +2\Delta_{12}\beta_1\cdot \beta_2}{\hat\alpha_1 \hat\alpha_2 - \Delta_{12}^2}}
\Biggr\rbrack^{1+\frac{3}{2}(D-4)}
\nonumber\\
\label{Gammafourkdone1}
\end{eqnarray}
Setting $D=4$, this becomes

\begin{eqnarray}
\Gamma (x,y,z_{1},z_{2},0,0) &=& \frac{4\lambda^2}{(4\pi)^6}
\int^{1}_{0} du_{1}du_2
\int_{-\bigtriangleup_{11}}^{\infty} d\hat\alpha_{1} 
\int_{-\bigtriangleup_{22}}^{\infty} d\hat\alpha_{2} 
\nonumber\\
&&\times 
\frac{1}{\Bigl[\hat\alpha_1 \hat\alpha_2 - \Delta_{12}^2\Bigr]
\Bigl[(x-y)^2(\hat\alpha_1 \hat\alpha_2 - \Delta_{12}^2) + \hat\alpha_1\beta_2^2 + \hat\alpha_2 \beta_1^2 +2\Delta_{12}\beta_1\cdot \beta_2\Bigr]
}
\, .
\nonumber\\
\label{Gammafourmassless}
\end{eqnarray}
Performing the $\hat\alpha_1$ - integral, which is elementary, we find

\begin{eqnarray}
\Gamma (x,y,z_{1},z_{2},0,0) &=& \frac{4\lambda^2}{(4\pi)^6}
\int^{1}_{0} du_{1}du_2
\int_{-\bigtriangleup_{22}}^{\infty} d\hat\alpha_{2} 
\nonumber\\
&&\times 
\frac{\ln \Biggl\lbrace
\frac{\hat\alpha_2 \bigl\lbrack \hat\alpha_2(\beta_1^2-\Delta_{11}(x-y)^2)+2\Delta_{12}\beta_1\cdot \beta_2-\Delta_{11}\beta_2^2-\Delta_{12}^2(x-y)^2\bigr\rbrack}
{(\hat\alpha_2(-\Delta_{11})-\Delta_{12}^2)(\hat\alpha_2(x-y)^2+\beta_2^2)}
\Biggr\rbrace
}
{(\hat\alpha_2\beta_1+\Delta_{12}\beta_2)^2}
\, .
\nonumber\\
\label{Gammafourmassless1}
\end{eqnarray}
The $\hat\alpha_2$ - integral is still a straightforward one. Introducing the 
zeroes $\hat\alpha_{\pm}$ of the quadratic form in the denominator,

\bear
\hat\alpha_{\pm} := -\frac{\Delta_{12}}{\beta_1^2}\Bigl\lbrack\beta_1\cdot\beta_2\pm i \sqrt{\beta_1^2\beta_2^2-(\beta_1\cdot\beta_2)^2}\Bigr\rbrack
\nonumber\\
\label{defhatalphapm}
\ear
we can write the result as

\begin{eqnarray}
\Gamma (x,y,z_{1},z_{2},0,0) &=& \frac{4\lambda^2}{(4\pi)^6}
\int^{1}_{0} du_{1}du_2
\frac{1}{(\hat\alpha_{+} - \hat\alpha_{-})\beta_1^2}
\biggl[
\ln \Biggl(
\frac{-\Delta_{11}a+\beta_1^2}
{-\Delta_{11}a}
\Biggr)
\ln \Biggl(\frac{-\Delta_{22}-\hat\alpha_-}{-\Delta_{22}-\hat\alpha_+}\Biggr)
\nonumber\\&&
+ I(0) + 
I\biggl(\frac{2\Delta_{12}\beta_1\cdot \beta_2-\Delta_{11}\beta_2^2-\Delta_{12}^2a}{\beta_1^2-\Delta_{11}a}\biggr)
- I\Bigl(\frac{\Delta_{12}^2}{\Delta_{11}}\Bigr) - I\Bigl(\frac{\beta_2^2}{a}\Bigr)
\biggr]
\nonumber\\
\label{Gammafourmassless2}
\end{eqnarray}
where

\bear
I(A) &:=& (\hat\alpha_{+} - \hat\alpha_{-})\int_{-\bigtriangleup_{22}}^{\infty} d\hat\alpha_{2} 
\frac{\ln (\hat\alpha_2 + A)}{(\hat\alpha_2 - \hat\alpha_{+})(\hat\alpha_2 - \hat\alpha_-)}
\nonumber\\
&=&
\biggl\lbrace
{\rm Li}_2\Bigl(\frac{A-\Delta_{22}}{A+\hat\alpha_-}\Bigr)
+ \ln (A-\Delta_{22})
\ln\Bigl(\frac{\hat\alpha_-+\Delta_{22}}{\hat\alpha_-+A}\Bigr)
\nonumber\\
&& +\frac{1}{2}\ln^2\Bigl(-\frac{1}{A+\hat\alpha_-}\Bigr)
\biggr\rbrace
- (\hat\alpha_- \to \hat\alpha_+)
\label{defI}
\ear
and we have abbreviated  $a : (x-y)^2$ as before. To rewrite the new integrand completely in terms
of the external Lorentz invariants, we further introduce

\begin{eqnarray}
b_i &:=& (x-z_i)^2 \, , \nonumber\\
c_i &:=& (y-z_i)^2 \, , \nonumber\\
d &:=& (z_1-z_2)^2 \, .
\label{defbicid}
\end{eqnarray}
In terms of these variables,

\begin{eqnarray}
 \beta_i^2 &=& u_i b_i + (1-u_i) c_i  - u_i(1-u_i)a \, ,\nonumber\\
2\beta_1\cdot\beta_2 &=& (2u_1u_2 -u_1 - u_2)a + u_2 b_1 + u_1 b_2 + (1-u_2)c_1
+ (1-u_1)c_2 - d \, . \nonumber\\
\label{elbeta}
\end{eqnarray}
Although we are not able to perform the remaining two integrals analytically, the representation (\ref{Gammafourmassless2}) is still more
explicit than other representations available for this integral which, as was mentioned in the introduction, plays an important role in
SYM theory \cite{36,bkpss,henhub,bogrpr}.

For the general $N$-rung case, the formulas (\ref{halfladder1}), (\ref{starting2}) generalize immediately to

\begin{eqnarray}
\Gamma (x,y,z_{1},z_{2},\ldots,z_{N}) &=& (-\lambda)^{N}\int^\infty_0 \frac{dT\, T^{N}}{(4\pi T)^{D/2}} e^{-\frac{(x - y)^2}{4T} -m^{2} T}
\int \frac{d^{D}k_1}{(2\pi)^D} \cdots \frac{d^{D}k_N}{(2\pi)^D} \, e^{-i\sum_{i=1}k_{i}\cdot z_{i}}
\nonumber\\
&&\times
\int d\alpha_{1}\cdots d\alpha_{N} e^{-\sum_{i=1}^{N}\alpha_{i}(k_{i}^2 + \mu^2)}\int du_{1}\ldots du_{N} 
e^{i\sum_{i=1}^{N}k_{i}\cdot(y+(x-y)u_{i})}
\nonumber\\
&&\times
{\rm exp}\Bigl[ T\sum_{i,j=1}^{N}\Delta_{ij}k_{i}\cdot k_{j} \Bigr]
\, .
\end{eqnarray}
The formulas (\ref{kdone1}), (\ref{Gammafourkdone}) generalize to

\begin{eqnarray}
\Gamma (x,y,z_{1},z_{2},\ldots,z_{N}) &=& 
\frac{(-\lambda)^{N}}{(4\pi)^{N\frac{D}{2}}}\int^\infty_0 \frac{dT\, T^{N(2-D/2)}}{(4\pi T)^{D/2}} e^{-\frac{(x - y)^2}{4T} -m^{2} T}
\int_0^1du_1\cdots du_N
\nonumber\\
&&\times
\int_0^{\infty} d\hat\alpha_{1}\cdots d\hat\alpha_{N}
e^{-\sum_{i=1}^N\hat\alpha_i \mu^2T}
\frac{1}{(\det H_N)^{\frac{D}{2}}} e^{-\frac{1}{4T}\vec b_N^{T}H_N^{-1}\vec b_N} \, .
\nonumber
\\
\end{eqnarray}
Here $H_N$ is the symmetric $N\times N$ matrix with entries

\bea
H_{Nii} &=& \hat\alpha_i  - \Delta_{ii} \, ,\nonumber\\
H_{Nij} &=& -\Delta_{ij} \quad (i\ne j) \, , \nonumber\\
\label{defHN}
\eea
and $\vec b_N = (\beta_1,\ldots,\beta_N)$ with $\beta_i$ as defined in (\ref{defbetai}).

Finally, also the massless four-dimensional formula (\ref{Gammafourmassless}) can still be generalized
to arbitrary $N$, in the form

\begin{eqnarray}
\Gamma (x,y,z_{1},z_{2},\ldots,z_N,0,0) &=& \frac{4(-\lambda)^N}{(4\pi)^{2(N+1)}}
\int^{1}_{0} du_{1}\cdots du_N
\int_0^{\infty} d\hat\alpha_{1} \cdots\hat\alpha_N
\nonumber\\
&&\times 
\frac{1}{(\det H_N)^2
\Bigl[(x-y)^2 + \vec b_N^{T}H_N^{-1}\vec b_N\Bigr]
}
\, .
\nonumber\\
\label{GammafourmasslessN}
\end{eqnarray}
It seems not to be possible, though, to do all the $\hat\alpha_i$ - integrals in closed form for general $N$.

\section{$N$-rung ladders}
\label{sec:ladders}
\renewcommand{\theequation}{4.\arabic{equation}}
\setcounter{equation}{0}

We will now come to our main purpose, namely to use the representations obtained for the $N$-propagators in
section \ref{sec:propagators} for constructing
the sum of all ladder and crossed-ladder graphs with $N$ rungs (simply called ``$N$-ladders'' in the following) in our
scalar Yukawa theory (\ref{S}).

Let us start with the graphs in
momentum space. Starting with the product of two copies of (\ref{orderNpspace}), identifying $k_i$ of one $N$ - propagator with
$-k_i$ of the second one, and inserting the connecting propagator integrals
$$\int \frac{dk_1}{(2\pi)^D}\frac{1}{ k_1^2+\mu^2} \ldots \int \frac{dk_N}{(2\pi)^D}\frac{1}{k_N^2+\mu^2}$$
produces precisely $N!$ times the $N$-ladder graphs (the momentum space versions of the graphs shown in fig. \ref{fig:ladders};
replace $y,{\bar y}, x, {\bar x}$ by (incoming) momenta $(p_1,p_2,q_1,q_2)$ there). We obtain the following integral representation
for the sum of these graphs:

\bea
\la\tilde \phi(q_1) \tilde \phi(q_2) \tilde \phi(p_1) \tilde \phi(p_2)\ra_{_{\!(N)}} 
&=&
(2\pi)^D \delta^D(p_1+p_2+q_1+q_2) 
 \frac{\lambda^{2N}}{N!} 
\int_0^{\infty}dS\int_0^{\infty}dT\,{\rm e}^{-m^2(S+T)}
\nonumber\\
&&\hspace{-0pt}\times\int_0^S d\sigma_1 \cdot\cdot\int_0^Sd\sigma_N\, 
\int_0^T d\tau_1 \cdot\cdot\int_0^Td\tau_N\,
\nonumber\\
&&\times
\int \frac{dk_1}{(2\pi)^D}\frac{1}{k_1^2+\mu^2} \ldots \int \frac{dk_N}{(2\pi)^D}\frac{1}{k_N^2+\mu^2}
(2\pi)^D \delta^D(p_1+p_2+ \sum_i k_i)
\nonumber\\
&& \times {\rm exp} \biggl\lbrace - S p_1^2 
- \sum_i (k_i^2+2p_1\cdot k_i )\sigma_i
- \sum_{i<j} 2k_i \cdot k_j D(\sigma_i,\sigma_j)\biggr\rbrace
\nonumber\\ &&\times
{\rm exp} \biggl\lbrace - T q_1^2 
- \sum_i (k_i^2-2q_1\cdot k_i )\tau_i
- \sum_{i<j} 2k_i \cdot k_j D(\tau_i,\tau_j)
\biggr\rbrace \, .
\nonumber\\
\label{ladderp1}
\eea
Next, we introduce Schwinger parameters $\alpha_1,\ldots,\alpha_N$ to exponentiate the
``rung'' propagators,
and we also (re-)exponentiate the second $\delta$ - function,

\bea
(2\pi)^D \delta^D\bigl(p_1+p_2+ \sum_i k_i\bigr) &=& \int dv\, {\rm e}^{iv\cdot \bigl(p_1+p_2+ \sum_i k_i\bigr)}
\, .
\label{expdelta}
\eea
The $k_i$ - integrals are now Gaussian, and performing them involves only the inverse and the determinant of
the symmetric $N\times N$ - matrix $A_N$ with entries 

\bea
A_{Nii} &=& \sigma_i + \tau_i + \alpha_i \, ,\nonumber\\
A_{Nij} &=& D(\sigma_i,\sigma_j)+ D(\tau_i,\tau_j)\quad (i\ne j)  \, .\nonumber\\
\label{defAN}
\eea
The $v$ - integral then also becomes Gaussian. Doing it one is left with the following integral
representation for the $N$ - ladder (henceforth we will omit the global $\delta$ function factor
$(2\pi)^D \delta^D(p_1+p_2+q_1+q_2)$): 

\bea
\la\tilde \phi(q_1) \tilde \phi(q_2) \tilde \phi(p_1) \tilde \phi(p_2)\ra_{_{\!(N)}} 
&=&
\frac{1}{(4\pi)^{(N-1)\frac{D}{2}}}
 \frac{\lambda^{2N}}{N!} 
\int_0^{\infty}dS\int_0^{\infty}dT\,{\rm e}^{-m^2(S+T)}
\nonumber\\
&&\hspace{-0pt}\times\int_0^S d\sigma_1 \cdot\cdot
\int_0^Td\tau_N\,
\int_0^{\infty} d\alpha_1 \cdot\cdot\int_0^{\infty}d\alpha_N\,
\frac{{\rm e}^{-\mu^2\sum_i \alpha_i}}{(a_N\,{\rm det}{A_N})^{\frac{D}{2}}}
\nonumber\\
&&
 \times 
 {\rm exp} \biggl\lbrace -Sp_1^2-Tq_1^2 - \frac{ b_N^2}{a_N}
 + (p_1\vec\sigma -q_1\vec\tau)\cdot A_N^{-1} \cdot (p_1\vec\sigma -q_1\vec\tau)
\biggr\rbrace
\, .
\nonumber\\
\label{ladderpfin}
\eea
Here we have further defined 

\bea
a_N &:=& \vec 1\cdot A^{-1}\cdot \vec 1\, , \nonumber\\
b_N &:=& p_1+p_2 - \vec 1\cdot A_N^{-1}\cdot \vec\sigma \, p_1+\vec 1\cdot A_N^{-1}\cdot \vec\tau \, q_1\, ,\nonumber\\
\label{defaNbN}
\eea
with $\vec 1 := (1,\ldots,1)$, $\vec\sigma:=(\sigma_1,\ldots,\sigma_N)$ etc. 
It is understood that the matrix $A_N$ acts trivially on Lorentz indices.
Note that (\ref{ladderpfin}) is still valid in $D$ dimensions.

Fourier transforming (\ref{ladderpfin}) we obtain the corresponding amplitude in $x$ - space
in the form

\bea
\la \phi_q(x)\phi_q(\bar x) \phi_q(y) \phi_q(\bar y) \ra_{_{\!(N)}} 
&=&
\frac{1}{(4\pi)^{(N+2)\frac{D}{2}}}
 \frac{\lambda^{2N}}{N!} 
\int_0^{\infty}dS\int_0^{\infty}dT\,{\rm e}^{-m^2(S+T)}
\nonumber\\
&&\hspace{-0pt}\times\int_0^S d\sigma_1 \cdot\cdot
\int_0^Td\tau_N\,
\int_0^{\infty} d\alpha_1 \cdot\cdot\int_0^{\infty}d\alpha_N\,
\frac{{\rm e}^{-\mu^2\sum_i \alpha_i}}{({\rm det}L{\rm det}{A_N})^{\frac{D}{2}}}
\nonumber\\
&&
 \times 
 {\rm exp} \biggl\lbrace - \frac{1}{4}\Bigl\lbrack a_N(y-\bar y)^2 + 
 (w,\bar w)L^{-1}(w,\bar w)
 \Bigr\rbrack
\biggr\rbrace
\nonumber\\
\label{ladderxfin1}
\eea
where 

\bea
L &:=&
\left(
\begin{array}{cc}
 S-\vec\sigma A_N^{-1}\vec\sigma &  \vec\sigma A_N^{-1}\vec\tau   \\
\vec\sigma A_N^{-1}\vec\tau  &  T-\vec\tau A_N^{-1}\vec\tau  \\ 
\end{array}
\right)
\, ,
\nonumber\\
w &:=& x-y + \vec 1 A_N^{-1}\vec\sigma \,(y-\bar y)\, , \nonumber\\
\bar w &:=& \bar x-\bar y - \vec 1 A_N^{-1}\vec\tau \,(y-\bar y)\, . \nonumber\\
\label{defBwwbar}
\eea
Starting instead directly from (\ref{orderNxspace}), one finds the alternative, very compact form

\bea
\la \phi_q(x)\phi_q(\bar x) \phi_q(y) \phi_q(\bar y) \ra_{_{\!(N)}} 
&=&
\frac{1}{(4\pi)^{(N+2)\frac{D}{2}}}
 \frac{\lambda^{2N}}{N!} 
\int_0^{\infty}dS\int_0^{\infty}dT\,{\rm e}^{-m^2(S+T)-\frac{(x-y)^2}{4S}-\frac{(\bar x-\bar y)^2}{4T}}
\nonumber\\
&&\hspace{-0pt}\times\int_0^S d\sigma_1 \cdot\cdot
\int_0^Td\tau_N\,
\int_0^{\infty} d\alpha_1 \cdot\cdot\int_0^{\infty}d\alpha_N\,
\frac{{\rm e}^{-\mu^2\sum_i \alpha_i - \frac{1}{4}\vec rM_N^{-1}\vec r}}{(ST{\rm det}{M_N})^{\frac{D}{2}}}
\nonumber\\
\label{ladderxfin2}
\eea
where $M_N$ is the symmetric $N\times N$ matrix 

\bea
M_{Nij} &:=& \delta_{ij}\alpha_i - \Delta_S(\sigma_i,\sigma_j) - \Delta_T(\tau_i,\tau_j)
 \nonumber\\
\label{defMN}
\eea
and

\bea
\vec r &:=&  (y-\bar y)\vec 1 +\frac{x-y}{S}\vec\sigma - \frac{(\bar x-\bar y)}{T}\vec\tau \, .
\label{defr}
\eea
We note that the two $x$-space representations (\ref{ladderxfin1}),(\ref{ladderxfin2}) can be related by 

\bea
M_N &=& A_N - \frac{\vec\sigma \otimes \vec\sigma}{S} -  \frac{\vec\tau \otimes \vec\tau}{T} \, ,
\nonumber\\
M_N^{-1} &=& A_N^{-1} + L_{11}^{-1}A_N^{-1}\cdot \vec\sigma\, \vec\sigma\cdot A_N^{-1} 
- L_{12}^{-1}A_N^{-1}\cdot \vec\sigma\, \vec\tau\cdot A_N^{-1} 
- L_{21}^{-1}A_N^{-1}\cdot \vec\tau\, \vec\sigma\cdot A_N^{-1} 
\nonumber\\
&& + L_{22}^{-1}A_N^{-1}\cdot \vec\tau\, \vec\tau\cdot A_N^{-1} \, ,
\nonumber\\\label{idMA}
\eea
which also implies that

\bea
ST{\rm det}M_N &=& {\rm det}L\, {\rm det}A_N \, .
\label{iddet}
\eea


\section{An application: lowest bound state mass from scalar ladders}
\label{sec:wc}
\renewcommand{\theequation}{5.\arabic{equation}}
\setcounter{equation}{0}

We proceed to the simplest possible application of our formulas for the ladder graphs to the physics of bound states,
which is the extraction of the lowest bound state mass.
Following \cite{nietjoPRL}, this can be done by considering the limit of large
timelike separation $t\to\infty$, where

\begin{eqnarray}
t &=& \frac{1}{2} (x_4+\bar x_4 - y_4 - \bar y_4)\, .
\label{deft}
\end{eqnarray}
Denoting the four -- point Green's function in the ladder approximation by $G$,

\begin{eqnarray}
G &=& \sum_{N=0}^{\infty} G_N = 
\sum_{N=0}^{\infty}\la \phi(x)\phi(\bar x) \phi(y) \phi(\bar y) \ra_{_{\!(N)}} 
\label{defG}
\end{eqnarray}
one has

\begin{eqnarray}
G &\stackrel{t\to\infty}{\simeq} & c_0\,{\rm e}^{-m_0t}
\label{m0fromG}
\end{eqnarray}
where $m_0$ is the lowest bound state mass. 
We can set $D=4$, since no regularization will be required.
Since we are not interested in the wave function of the bound state at present, we can simplify the formula for $G$ by setting

\begin{eqnarray}
x = \bar x, \quad y = \bar y
\label{simp}
\end{eqnarray}
so that $t= x_4 - y_4= \bar x_4 - \bar y_4$. Further, since the limit $t\to\infty$ is taken at finite spatial displacement,
in this limit we can effectively set 

\begin{eqnarray}
t^2 &= & (x-y)^2 = (\bar x - \bar y)^2 \, .
\label{simpt}
\end{eqnarray}
Using these relations in eqs. (\ref{ladderxfin2}), introducing the dimensionless time parameter

\begin{eqnarray}
\hat t := \frac{m}{2}t
\label{defthat}
\end{eqnarray}
as well as the effective coupling constant

\begin{eqnarray}
g &:= & \frac{\lambda^2}{(4\pi)^2m^2} \, ,
\label{defg}
\end{eqnarray}
rescaling $S,T,\alpha_i$ all by a factor $\hat t/m^2$, and changing variables from $\sigma_i,\tau_i$ to
$u_i,v_i$ through

\bear
\sigma_i = S\frac{\hat t}{m^2}u_i, \quad \tau_i = T\frac{\hat t}{m^2}v_i \, ,
\label{elsigmatau}
\ear
we get our following ``master formula'',

\bea
G_N
&=&
\frac{m^4}{(4\pi)^4\hat t^2}
 \frac{(\hat t g)^N}{N!} 
\int_0^{\infty}dS\,S^{N-2}\int_0^{\infty}dT\,T^{N-2}
\nonumber\\
&&\hspace{-0pt}\times\int_0^1 d u_1 \cdot\cdot
\int_0^1d v_N\,
\int_0^{\infty} d\alpha_1 \cdot\cdot\int_0^{\infty}d\alpha_N\,
\frac{1}{({\rm det}{\hat M_N})^2}
\nonumber\\
&&
 \times 
 {\rm exp} \biggl\lbrace - \hat t
 \biggl\lbrack
S + T +  \frac{1}{S} + \frac{1}{T} + \frac{\mu^2}{m^2}\sum_i \alpha_i
+ (\vec u -\vec v)\hat M_N^{-1} (\vec u-\vec v )
\biggr\rbrack\biggr\rbrace
\nonumber\\
\label{GNmaster}
\eea
where now

\bea
\hat M_{Nij} &:=& \delta_{ij}\alpha_i - S\Delta_1(u_i,u_j) - T\Delta_1(v_i,v_j) \, .
 \nonumber\\
\label{defhatMN}
\eea
We remark that in \cite{barroberg:fbs}, inspired
by Feynman's famous treatment of the polaron problem \cite{feynman-polaron}, Barro-Bergfl\"odt, Rosenfelder 
and Stingl have approximated the action in the worldline or Feynman-Schwinger 
representation of the four-point Green's function $G$ (but including the
self-energy and vertex corrections) by a quadratic trial action, in order to obtain an
approximate value for the mass of the lowest-lying bound state.
Here, we will use the large $\hat t$ limit to eliminate, at fixed $S,T,\alpha_i, u_i$,
the $v_i$ integrals by a Gaussian approximation around the point $\vec v = \vec u$. 
For the validity of this approximation, it is essential that the matrix $\hat M_N$ be positive semidefinite, 
which we have checked numerically for various values of $N$.
The Gaussian approximation results in

\bea
G_N
&=&
\frac{m^4}{(4\pi)^4\hat t^2}
 \frac{(\pi\hat t g^2)^{N/2}}{N!} 
\int_0^{\infty}dS\,S^{N-2}\int_0^{\infty}dT\,T^{N-2}
\nonumber\\
&&\hspace{-0pt}\times\int_0^1 d u_1 \cdot\cdot
\int_0^1d u_N\,
\int_0^{\infty} d\alpha_1 \cdot\cdot\int_0^{\infty}d\alpha_N\,
\frac{1}{({\rm det}{\bar M_N})^{3/2}}
\nonumber\\
&&
 \times 
 {\rm exp} \biggl\lbrace - \hat t
 \biggl\lbrack
S + T +  \frac{1}{S} + \frac{1}{T} + \frac{\mu^2}{m^2}\sum_i \alpha_i
\biggr\rbrack\biggr\rbrace
\nonumber\\
\label{GNmaster1}
\eea
where now

\bea
\bar M_{Nij} &:=& \delta_{ij}\alpha_i - (S+T)\Delta_1(u_i,u_j) \, .
 \nonumber\\
\label{defbarMN}
\eea
After a further rescaling 

\bea
\alpha_i \equiv (S+T)\hat \alpha_i
\label{rescale1}
\eea
and summation over $N$, we obtain the following representation for the
full Green's function:

\bea
G
&=&
\frac{m^4}{(4\pi)^4\hat t^2}
\int_0^{\infty}\frac{dS}{S^2}\int_0^{\infty}\frac{dT}{T^2}
\, {\rm exp} \biggl\lbrace - \hat t
 \biggl\lbrack
S + T +  \frac{1}{S} + \frac{1}{T} 
\biggr\rbrack\biggr\rbrace
\nonumber\\
&&\times
\sum_{N=1}^{\infty} 
 \frac{(\pi\hat t g^2)^{N/2}}{N!} 
\biggl\lbrack\frac{ST}{(S+T)^{1/2}}\biggr\rbrack^N
c_N\Bigl(\hat t (S+T)\mu^2 /m^2\Bigr)
\nonumber\\
\label{GNmaster2}
\eea
where 
\bea
c_N(x) := \int_0^1 du_1 \cdots \int_0^1d u_N\,
\int_0^{\infty} d\hat \alpha_1 \cdot\cdot\int_0^{\infty}d\hat \alpha_N\,
\frac{{\rm e}^{-x\sum_i \hat \alpha_i}}{({\rm det}{H_N})^{3/2}}
\label{defcN}
\eea
and the matrix $H_N$ had already been introduced in (\ref{defHN}), 

\bea
H_{Nij} &=& \delta_{ij}\hat \alpha_i - \Delta_1(u_i,u_j) \, .
\eea
It should be noted that, in diagrammatic terms, our Gaussian approximation 
$\vec v = \vec u$ corresponds to proper ladder graphs. 
The only case where a trace of the crossed ladder graphs can still be left over is for
``overlapping rungs'' $v_i = u_i = v_j = u_j$ ($i \neq j$) which can be
obtained as limits of crossed or uncrossed rungs. 

We will determine the large-$\hat t$ behavior of $G$ (in a special case) by using a saddle point approximation in the representation
(\ref{GNmaster2}). 
First, however, we have to focus our attention on the functions $c_N$.
The integrals in (\ref{defcN}) are convergent, however this is not very  
transparent the way they are written. This motivates the following transformations. 
To begin with, let us rewrite the matrix $H_N$ as
\bea
H_N = D_N(\Eins-R_N)
\label{defM}
\eea
\\
where $D_N$ is the diagonal part of $H_N$
\bea
D_{Nij}&:=&\delta_{ij}(\hat \alpha_i-\Delta_1(u_i,u_j))=\delta_{ij}(\hat \alpha_i+u_i(1-u_i))
\label{defDiag}
\eea
and
\bea
R_N:=D_N^{-1}\Delta^{'} 
\label{defR}
\eea
where $\Delta'$ denotes the matrix $\Delta_{ij}$ with  its diagonal terms deleted (here we use the abbreviated notation $\Delta_{ij} = \Delta_1 (u_i, u_j)$, as before).
Then, we perform a change of variables from $\hat \alpha_i$ to $\hat \beta_i$
\bea
\hat \beta_i:=\sqrt\frac{-\Delta_{ii}}{\hat \alpha_i-\Delta_{ii}} \, .
\eea
The integrals in (\ref{defcN}) then turn into
\bea
c_N(x) = 2^N\int_0^1\frac{du_1}{\sqrt{u_1(1-u_1)}}\cdots\int_0^1\frac{du_N}{\sqrt{u_N(1-u_N)}} \int_0^1d\hat \beta_1\cdots\int_0^1 d\hat \beta_N  \frac{{\rm exp}^{-x\sum_i(-\Delta_{ii})(\frac{1}{\hat \beta_i^2}-1)}}{{\rm det}^{3\over2}(\Eins-R)}
\, .
\nonumber\\
\label{defcn}\eea
Note that now $D_{Nij}^{-1} = \delta_{ij}\hat \beta_i^2/(-\Delta_{ii})$.

Further, since the integrand is permutation symmetric, the full $u_i$ integrals can be replaced by $N!$ times the integral over the ordered sector $u_1\ge u_2\ge u_3 \cdots \ge u_N$. Thus we define
\\
\bea
\bar c_N(x) := \frac{c_N(x)}{2^{N}N!} = \int_0^1\frac{du_1}{\sqrt{u_1(1-u_1)}}\int_0^{u_1}\frac{du_2}{\sqrt{u_2(1-u_2)}}\cdots\int_0^{u_{N-1}}\frac{du_N}{\sqrt{u_N(1-u_N)}}
\nonumber \\
\times\int_0^1d\hat \beta_1\cdots\int_0^1 d\hat \beta_N \frac{{\rm exp}^{-x\sum_i(-\Delta_{ii})(\frac{1}{\hat \beta_i^2}-1)}}{\rm det^{3\over2}(\Eins-R)}
\, .
\nonumber\\
\label{defcnbar}
\eea

From now on, we will focus on the case of a massless particle exchange $\mu = 0$, where the functions $\bar c_N(x)$ reduce to numbers

\bea
\bar c_N(0)=:\bar c_N \, .
\eea
The first coefficient is 

\bea
\bar c_1 = \int_0^1\frac{du_1}{\sqrt {u_1(1-u_1)}}=\pi \, .
\label{c1}
\eea
For $N>1$, inspection of the determinant $\rm det(\Eins -R)$ shows that
it simplifies considerably if, instead of $u_1,\ldots,u_N$, one writes  it in terms of new variables $z_2,\ldots,z_N$ defined by
conformal cross ratios,

\bea
z_i:=\sqrt{\frac{u_i(1-u_{i-1})}{u_{i-1}(1-u_i)}} \, .
\eea
Changing variables from $u_i$ to $z_i$ for $i=2$,$\ldots$,$N$, we obtain

\bea
\bar c_N = 2^{N-1}\int_0^1dz_2\int_0^1dz_3\cdots\int dz_N {\cal M}_N\int_0^1 d\hat \beta_1 \cdots \int_0^1 d\hat \beta_N \frac{1}{\rm det^{3\over2}(\Eins-R)}
\nonumber\\
\label{CN}
\eea
where R is now written as a function of $\hat \beta_1,\ldots,\hat \beta_N, z_2,\ldots, z_N$ and ${\cal M}_N$ is a function of $z_2,\ldots,z_N$ defined as

\bea
{\cal M}_N:=\frac{1}{z_2z_3\cdots z_N}\int_0^1du_1\sqrt{\frac{u_2(1-u_2)u_3(1-u_3)\cdots u_N(1-u_N)}{u_1(1-u_1)}}
\nonumber\\
\label{MN}
\eea
Here it is understood that first $u_2,\ldots,u_N$ are, backwards starting from $u_N$, transformed to $z_2,\ldots,z_N$ via

\bea
u_i=\frac{u_{i-1}z_i^2}{1-u_{i-1}(1-z_i^2)}
\label{ui}
\eea
\\
$(i\ge 2)$ and then the $u_1$ integral is performed. For $N=2,3,$ one finds
\\
\bea
{\cal M}_2=\frac{2\log z_2}{(z_2-1)(z_2+1)} \, ,
\eea
\bea
{\cal M}_3=\frac{\pi}{(z_2+1)(z_3 +1)(z_2z_3+1)} \, .
\eea
After this transformation, the integral for the second coefficient, too, has become elementary: 

\bea
\bar c_2 = 2\int_0^1dz_2 {\cal M}_2 \int_0^1 d\hat \beta_1\int_0^1d\hat \beta_2\frac{1}{(1-\hat \beta_1^2\hat \beta_2^2z_2^2)^\frac{3}{2}} =\frac{\pi^3}{6}
\, .
\eea
The next coefficients up to $N=11$ could be determined by numerical integration employing
the representation (\ref{CN}), see table \ref{t1}.

\bigskip

\begin{table*}[h]
\noindent
\caption[t1]{The coefficients $\bar c_N$.} 

\bigskip

\begin{tabular}{|c|ccccccccccc|}
\hline
$N$ & 1 & 2 & 3 & 4 & 5 & 6 & 7 & 8 & 9 & 10& 11  \\
\hline
$\bar c_N$ & $\pi$ & $\frac{\pi^3}{6}$  & $5.9319$  & $5.3402$ & $4.0192$ & $2.6243$ & $1.5349$ & $0.8044$ & $0.378$ & $0.175$ & $0.076$   \\
\hline
\end{tabular}
\label{t1}
\end{table*}

Since an exact calculation of these coefficients for general $N$ seems out of the question, we will now try to determine their
asymptotic behaviour in the large-$N$ limit. 
We begin by asking what
the asymptotic behavior of the coefficients $\bar c_N$ {\it should} be to get the expected
correction to the lowest bound state mass in the nonrelativistic limit. In this limit, the exact bound state energy
would, for $\m=0$, be \cite{salbet:pr84,itzzub-book}

\bea
E_b = \frac{1}{4}m\alpha^2
\label{Eb}
\eea
where 

\bea
\alpha = \frac{\lambda^2}{16 \pi m^2} = \pi g \, .
\label{defalpha}
\eea
This corresponds to an exponential factor

\bea
{\rm e}^{-Et} = {\rm e}^{-(2m-E_b)t} = {\rm e}^{-(2m-E_b)2\hat t/m} = {\rm e}^{(-4+\frac{1}{2} \pi^2g^2)\hat t}
\label{explarget}
\eea
for the large-$\hat t$ behavior of $G$.
This should become the exact answer for small $g$.
Now,  in the representation (\ref{GNmaster2}) of $G$ the trivial exponent $-4\hat t$ corresponds to a saddle point at
$S=T=1$; thus, at least for small $g$ it should be a good approximation to set $S=T=1$
also in the factor $[ST/(S+T)^{1/2}]^N$
that appears in the sum over $N$. This leaves us with the
series (cf.\ eq.\ (\ref{GNmaster2}) with $\mu = 0$)

\bea
\sum_N \frac{c_N}{N!}  \Bigl(\frac{\pi \hat t}{2}\Bigr)^{N/2} g^N
=  \sum_N \bar c_N  \bigl(2\pi \hat t \bigr)^{N/2} g^N \quad
\stackrel{!}{=}\quad 
 {\rm e}^{\frac{1}{2} \pi^2g^2\hat t} 
 \, .
\label{wt}
\eea
From the Taylor series

\bea
\sum_{N=0}^{\infty} \frac{x^N}{\Gamma (1+N/2)} = (1+{\rm Erf}(x))\,{\rm e}^{x^2} \stackrel{x\to\infty}{\sim} 2 \,{\rm e}^{x^2}
\label{taylor}
\eea
we then conclude that the $\bar c_N$ should have the asymptotic behavior

\bea
\bar c_N  \stackrel{N\to\infty}{\sim} \frac{c_{\infty}\beta^N}{\Gamma (1+N/2)} 
\label{wt2} 
\eea
which would lead to an exponential

\bea
{\rm e}^{2\pi\beta^2g^2\hat t} \, .
\label{betaexp}
\eea
Comparison with (\ref{wt}) yields

\bea
\beta \stackrel{!}{=} \frac{\sqrt{\pi}}{2} = 0.886 \, .
\label{limbeta}
\eea
To compare with our numerical results for the $\bar c_N$, we note that 
from (\ref{wt2}) it follows that the sequence

\bea
\beta_N := \frac{\bar c_{N+1}\Gamma\bigl(1+\frac{N+1}{2}\bigr)}{\bar c_{N}\Gamma\bigl(1+\frac{N}{2}\bigr)}
\label{defgamma_N}
\eea
should converge to $\beta$ for $N\to\infty$. The values of $\beta_N$ for
$N$ from $1$ to $10$ are given in table \ref{t2}, using the numerical values
for the coefficients $\bar c_N$ from table \ref{t1}.

\bigskip

\begin{table*}[h]
\noindent
\caption[t1]{The coefficients $\beta_N$.} 

\bigskip

\begin{tabular}{|c|cccccccccc|}
\hline
$N$ & 1 & 2 & 3 & 4 & 5 & 6 & 7 & 8 & 9 & 10 \\
\hline
$\beta_N$ & $1.856$ & $1.525$  & $1.355$  & $1.251$ & $1.179$ & $1.134$ & $1.081$ & $1.025$ & $1.061$ & $1.043$ \\
\hline
\end{tabular}
\label{t2}
\end{table*}

\begin{figure}[h]
\centering
\includegraphics{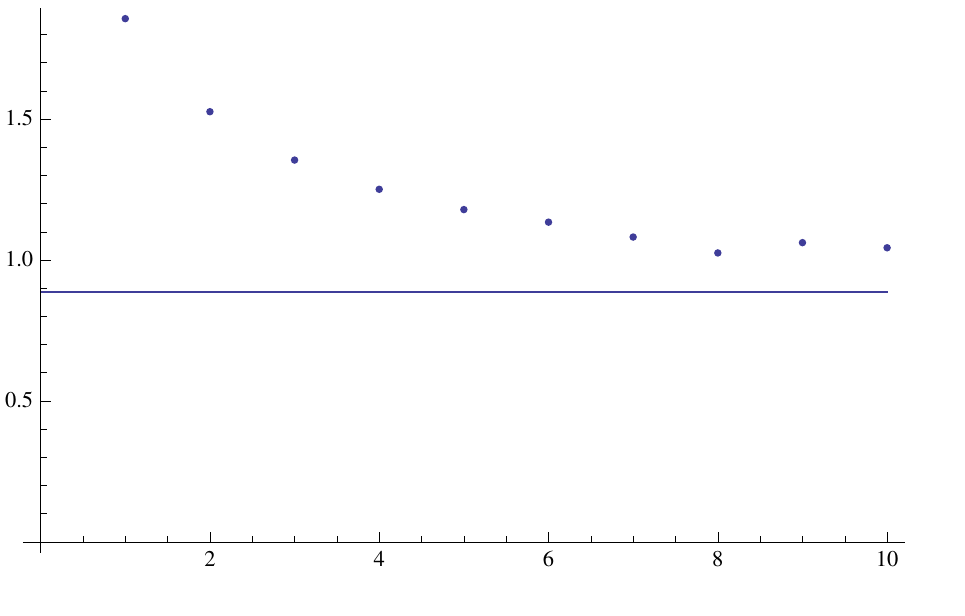}
\caption{The coefficients $\beta_N$}
\label{figbeta}
\end{figure}

\noindent
We have also plotted the $\beta_N$ together with the expected asymptotic limit $\beta$ in fig. \ref{figbeta}.

\noindent
The plot suggests that, if there is convergence at all, it will be to a higher value than $\beta$.

In order to understand what is going on, let us return to the coefficients $\bar c_N$ of table \ref{t1}, and plot the
combination

\begin{eqnarray}
\tilde c_N :=  \Gamma \left(1+ \frac{N}{2}\right)\frac{\bar c_N}{\beta^N} \, .
\label{deftildecn}
\end{eqnarray}

\begin{figure}[h]
\centering
\includegraphics{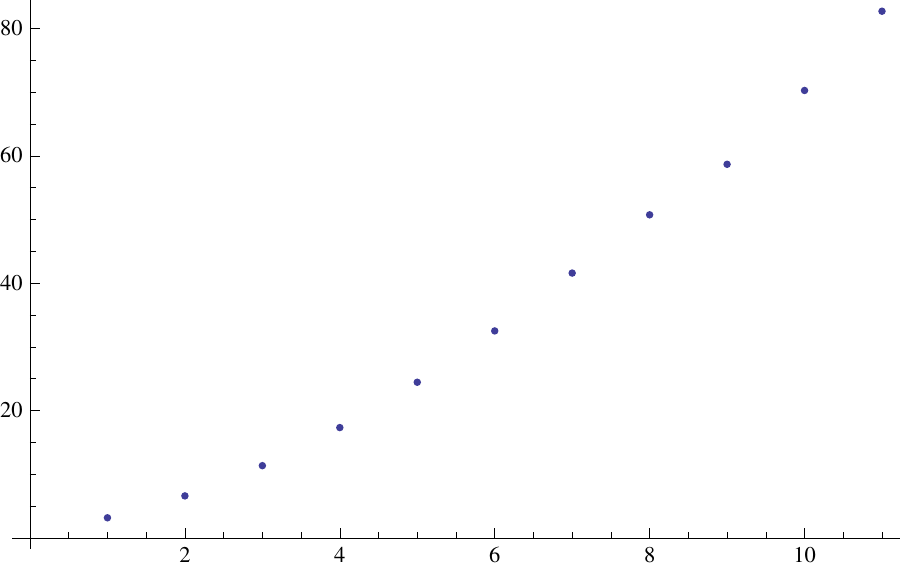}
\caption{The coefficients $\tilde c_N$}
\label{figctilde}
\end{figure}

\noindent
If (\ref{wt2}) were true, the coefficients would converge to the constant $c_{\infty}$; instead we find a curve which looks parabolic, see fig. \ref{figctilde}.

\noindent
Therefore, let us look at yet another set of coefficients $c_N'$,

\begin{eqnarray}
c_N' :=  \frac{\tilde c_N}{N^2} \, .
\label{defcnprime}
\end{eqnarray}

\begin{figure}[h]
\centering
\includegraphics{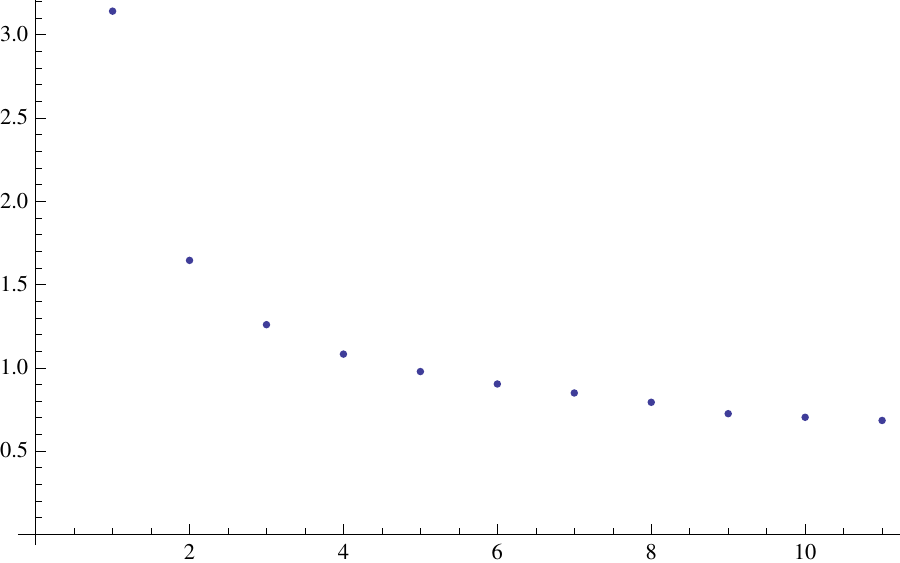}
\caption{The coefficients $c_N'$}
\label{figcnprime}
\end{figure} 

\noindent
These modified coefficients indeed seem to converge to a constant (see fig. \ref{figcnprime});  let us call this constant $c_{\infty}'$.
Thus we now have, instead of (\ref{wt2}), the asymptotic behaviour

\bea
\bar c_N  \stackrel{N\to\infty}{\sim} \frac{c_{\infty}'N^2\beta^N}{\Gamma (1+N/2)} \, .
\label{wt2prime} 
\eea
Fortunately, this does not change anything essential: instead of (\ref{taylor}) we get

\bea
\sum_{N=0}^{\infty}N^2 \frac{x^N}{\Gamma (1+N/2)}  \stackrel{x\to\infty}{\sim} 8x^4 \,{\rm e}^{x^2} \, .
\label{taylorprime}
\eea
So, there is no modification of the exponent, only of the prefactor, which does not interest us right now.\footnote{It is curious to note, however, 
that this change of the prefactor precisely removes the factor $1/\hat t^2$ in the master formula (\ref{GNmaster2}).}
We can also adapt the definition (\ref{defgamma_N}) of $\beta_N$ to the asymptotic behavior
(\ref{wt2prime}) by defining

\bea
\beta_N' := \frac{N^2 \bar c_{N+1}\Gamma\bigl(1+\frac{N+1}{2}\bigr)}{(N+ 1)^2 \bar c_{N}\Gamma\bigl(1+\frac{N}{2}\bigr)}
= \frac{\beta_N}{(1 + 1/N)^2} \, .
\label{defgammapr_N}
\eea
The first ten coefficients $\beta_N'$ are given in table \ref{t3}.

\bigskip

\begin{table*}[h]
\noindent
\caption[t1]{The coefficients $\beta_N'$.} 

\bigskip

\begin{tabular}{|c|cccccccccc|}
\hline
$N$ & 1 & 2 & 3 & 4 & 5 & 6 & 7 & 8 & 9 & 10 \\
\hline
$\beta_N'$ & $0.464$ & $0.678$ & $0.762$ & $0.801$  & $0.818$ & $0.833$ & $0.827$ & $0.817$ & $0.859$ & $0.861$ \\
\hline
\end{tabular}
\label{t3}
\end{table*}

\begin{figure}[h]
\centering
\includegraphics{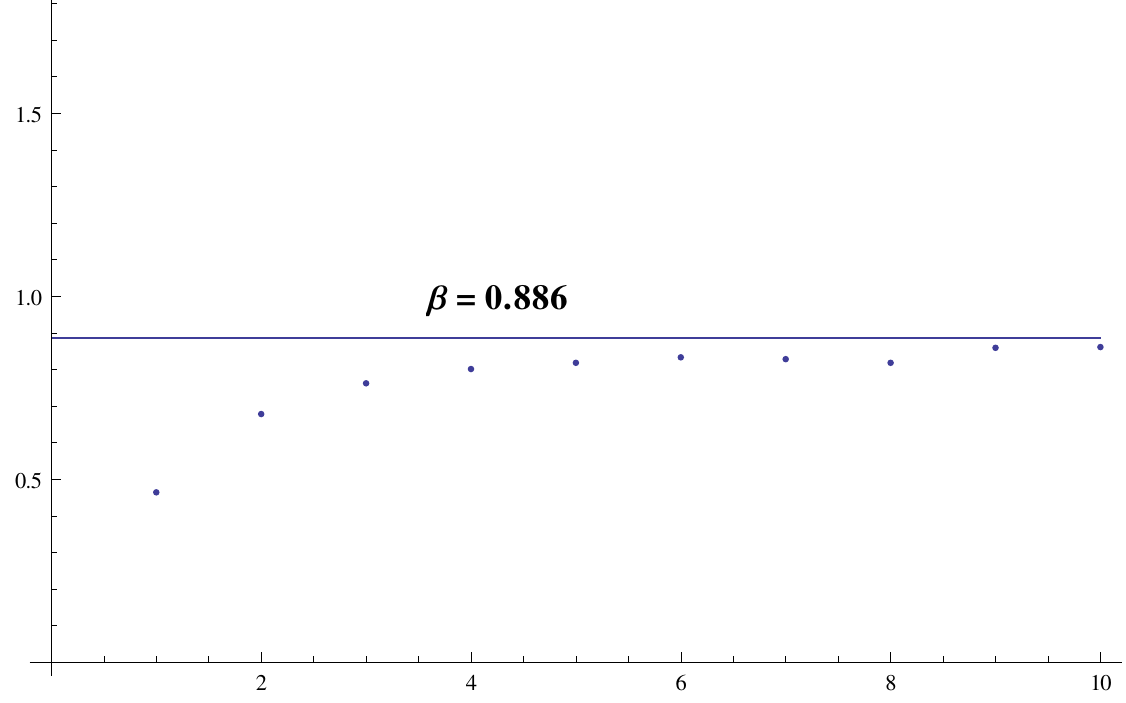}
\caption{The coefficients $\beta_N'$}
\label{fig:bnprime}
\end{figure}

\noindent
From these values, it is at least credible that $\beta_N'$ asymptotically converges to $\beta = 0.886$; see fig. \ref{fig:bnprime}.

\noindent

In the following, we hence assume
that (\ref{wt2prime}) is true, with $\beta = \sqrt{\pi}/2$. Let us then undo the assumption of small $g$ and of the saddle point at $S=T=1$
and return to (\ref{GNmaster2}). The asymptotic summation formula (\ref{taylorprime}) now leads to a total exponential factor

\bea
{\rm exp}\biggl\lbrack -\hat t
\Bigl(
S+T + \frac{1}{S} + \frac{1}{T} - \pi^2g^2 \frac{S^2T^2}{S+T} 
\Bigr)\biggr\rbrack \, .
\eea
As long as $g^2 < 1/3 \pi^2$, one finds a saddle point (local
maximum) of the exponent at

\bea
S=T= \sqrt{\frac{2}{3}}\frac{1}{\pi g}\sqrt{1-\sqrt{1-3\pi^2 g^2}}
\label{saddle}
\eea
with saddle point value

\bea
{\rm exp}\biggl\lbrace -\hat t
\frac{4 \sqrt{2}}{3} \biggl\lbrack 
\bigg ( 1 + \sqrt{1 - 3 \pi^2 g^2} \bigg)^{-1/2} +
\bigg ( 1 + \sqrt{1 - 3 \pi^2 g^2} \bigg)^{1/2} \biggr\rbrack
\biggr\rbrace \, .
\eea
From (\ref{m0fromG}), (\ref{defthat}) this gives for the lowest bound state mass $m_0$

\bear
\frac{m_0}{m} &=& 
\frac{2 \sqrt{2}}{3} \biggl\lbrack 
\bigg ( 1 + \sqrt{1 - 3 \pi^2 g^2} \bigg)^{-1/2} +
\bigg ( 1 + \sqrt{1 - 3 \pi^2 g^2} \bigg)^{1/2} \biggr\rbrack
\nonumber\\
\label{m0fin}
\ear
As $g^2$ increases from zero to its maximal value $1/3 \pi^2$, the
result (\ref{m0fin}) for this mass $m_0$ decreases 
monotonically from $2m$ to $\frac{4\sqrt{2}}{3} m = 1.886 m$. 
An expansion of (\ref{m0fin}) in $g$ yields 

\bea
\frac{m_0}{m} &=& 
2 - \frac{\pi^2g^2}{4} - \frac{9}{64}(\pi^2g^2)^2- \frac{81}{512}(\pi^2g^2)^3 - \ldots
\label{m0finexpg}
\ear
In the second term of the expansion we find again, of course, the nonrelativistic limit (\ref{Eb}) of the binding energy,
which we have already used as an input for our matching procedure; 
but the order $g^4$ term is already new. We note that in the
expansion (\ref{m0finexpg}) of the bound state mass in powers of $g$ no term of the
order $g^3 \ln g$ appears, as it would be the case for the corresponding
result in the Wick-Cutkosky model, \hbox{\it i.e.}, for the ladder
approximation of the Bethe-Salpeter equation in the same model theory
\cite{felful:prd7}. As we have mentioned before in the introduction, such a 
contribution is generally considered to be unphysical.

Our result for the mass of the lowest bound state may be compared
to the result of the relativistic eikonal approximation or Todorov's equation
\cite{brezitz:prd1,todorov:prd3}, in our notation,
\bear
\frac{m_0}{m} &=& 
\sqrt{2} \bigg ( 1 + \sqrt{1 - \pi^2 g^2} \bigg)^{1/2} \nonumber \\
&=& 2 - \frac{\pi^2g^2}{4} - \frac{5}{64}(\pi^2g^2)^2- \ldots
\label{todorov}
\ear
In terms of diagrams, the eikonal approximation sums up
all ladder and crossed ladder diagrams, but neglects any self-energy
contributions and vertex corrections, just as our approach does. It has been
argued to reproduce the contribution of the ladder and
crossed ladder diagrams correctly up to the order $g^4$ (see, \textit{e.g.}, \cite{barroberg:fbs}).
Note that the coefficient of the $g^4$-term in the expansion (\ref{todorov}) of the
bound state mass in powers of the coupling constant is somewhat smaller (in absolute value)
than in our approximation, but it has the same sign.

Finally, we can compare the maximal possible value of the
coupling constant, $g^2 = 1/3 \pi^2$, to the critical value found in the variational worldline approximation of \cite{barroberg:fbs}.
The latter value is (approximately) $\alpha = 0{.}814$ (without self-energy
and vertex corrections, and for a massless exchanged particle), somewhat larger than our value $\alpha = \pi g = 1/\sqrt{3} = 0{.}577$.
The existence of a critical coupling constant is attributed to the instability
of the vacuum in the scalar model theory in \cite{barroberg:fbs}.

\section{Conclusions}
\label{sec:conclusions}
\renewcommand{\theequation}{6.\arabic{equation}}
\setcounter{equation}{0}

To summarize, in this paper we have used the worldline formalism to derive integral representations for three classes of 
amplitudes - the $N$ - propagators, $N$ - half-ladders and the $N$ -  ladders - in scalar field theory involving an exchange of $N$ momenta, and 
in each case have given a compact expression combining the $N!$ Feynman diagrams contributing to the amplitude. 
For the $N$ - propagators and $N$ - ladders we have given these representations in both $x$ and (off-shell) momentum space,
for the $N$ - half-ladders in $x$ - space only. 
These amplitudes are not only of interest in their own right, but, being off-shell, can also been used as
building blocks for many more complex amplitudes. 

In particular, we have derived a compact expression for the
sum of all ladder graphs with $N$ rungs, including all possible crossings
of the rungs, which we use in section \ref{sec:wc} to extract an approximate
formula for the mass of the lowest-lying bound state, explicitly for the
case of a massless particle exchange between the constituents. Technically,
we apply a saddle point approximation to our formula for the $N$-rung
ladders, after summing over all $N$. {\em Before} applying the saddle point
approximation, however, we have made use of a Gaussian approximation in
eq.\ (\ref{GNmaster1}) that leads to an important simplification in the formulas for
the $N$-rung ladders. Both approximations exploit the large-time limit
that is being considered for the extraction of the lowest-lying bound
state, but it would certainly be more satisfying to have a way to arrive
at an approximate formula for the lowest bound-state mass by taking
advantage of the large-time limit in a single step, instead of using two
consecutive approximations.  Thus our procedure cannot claim mathematical rigor, 
but we think it is worth presenting it in any case. 
This is because, differently from previous attempts at this calculation \cite{wick,cutkosky,felful:prd7,itzzub-book},
in our approach  the truncation to the non-crossed ladder graphs is induced naturally by the
Gaussian approximation $\vec v = \vec u$, rather than done {\it ad hoc} from
the beginning, and moreover our final result (\ref{m0fin}) 
for the mass of the lowest bound state does not  display any obvious
inconsistencies. Equation (\ref{m0fin}) is similar
to the result of the relativistic eikonal approximation \cite{brezitz:prd1,todorov:prd3},
and the maximal value of the coupling constant for which a bound state
is found in our approximation is comparable to the critical coupling constant
in a variational worldline approximation \cite{barroberg:fbs}.
We intend to further test our result by a direct numerical path integral
calculation along the lines of  \cite{simtjo:ap228}, but taking advantage of the 
sophisticated worldline Monte Carlo technology developed in the meantime in \cite{gielan,giekli}.
Our aim in the present paper has merely been to demonstrate 
the feasibility of extracting information on the bound states of a theory 
from an analytical evaluation of the worldline
integrals, in an appropriate approximation.

Our second nontrivial application was to
obtain a new two-parameter integral representation for a 
massless four-point $x$ - space integral of some importance
in $N=4$ SYM theory  \cite{36,bkpss,henhub,bogrpr}. 

Coming to possible generalizations, it would be straightforward to extend our various
master formulas to the case of scalar QED (i.e. scalar lines and photon exchanges). 
In the spinor QED case (fermion lines and photon exchanges) closed-form expressions
for general $N$ could still be achieved using the worldline super-formalism \cite{41},
however at the cost of introducing additional multiple Grassmann integrals.
For eventual extensions to the nonabelian case it may turn out essential to work with a
path integral representation of the color degree of freedom, such as the one recently given in 
\cite{Bastianelli:2013pta}, rather than with explicit color factors.
Finally, even a closed-form treatment of ladder graphs involving the exchange of gravitons
between scalars or spinors - a completely hopeless task in the Feynman diagram approach
due to the existence of vertices involving an arbitrary number of gravitons - may be feasible
in the worldline formalism along the lines of \cite{Bastianelli:2002fv,Bastianelli:2013tsa}.


%
%
%
%
%
%

\bigskip

\noindent
{\bf Acknowledgements:} We would like to thank A. Davydychev, J. Henn and D.G.C. McKeon
for discussions and correspondence. C.S. thanks D. Kreimer and the Mathematical Physics group
of HUB for hospitality and discussions, as well as the HUB Gruppe Rechentechnik for access to their
supercomputing facility. A.H., C.S. and R.T. thank CONACyT for financial support.
A.W. acknowledges support by CIC-UMSNH and CONACyT project no.\
CB-2009/131787.

\begin{appendix}

\section{Comparison with Feynman diagrams}
\label{sec:feynman}
\renewcommand{\theequation}{A.\arabic{equation}}
\setcounter{equation}{0}

Let us consider the term appearing in (\ref{orderNpspace})
\bea
N!\int_0^1du_1 \cdot\cdot\int_0^1du_N\, 
\bigg [p_1^2 +m^2 + \sum_i 
(k_i^2+2p_1\cdot k_i )u_i
+ \sum_{i<j} 2k_i \cdot k_j 
\big ( u_i \theta(u_j -u_i)+ 
u_j \theta(u_i -u_j)\big )\bigg]^{-N-1} 
\eea
The integration region can be split into $N!$ subregions
specified by a unique ordering $\sigma(i)$ of the indices $i=1,2,..,N$
so that  $t_i=u_{\sigma(i)}$ are ordered as
$1\geq t_1 \geq t_2\geq ...\geq t_N \geq 0$.
Then each integration subregion contributes 
\bea
&& N!\int_0^1dt_1  \int_0^{t_1} dt_2 \int_0^{t_2} dt_3 
\cdot\cdot \int_0^{t_{N-1}}  dt_N\, 
\bigg [p_1^2 +m^2 + \sum_i 
(k_{\sigma(i)}^2 
+2 k_{\sigma(i)} \cdot p_1 )t_i \nonumber
\\
&& \hskip 3cm 
+ \sum_{i<j} 2k_{\sigma(i)} \cdot k_{\sigma(j)} 
\big ( t_i \underbrace{\theta(t_j -t_i)}_{=0}
+ t_j \underbrace{\theta(t_i -t_j)}_{=1}
\big )\bigg]^{-N-1} \nonumber
\\
&& = N!\int_0^1dt_1  \int_0^{t_1} dt_2 \int_0^{t_2} dt_3 
\cdot\cdot \int_0^{t_{N-1}}  dt_N\, 
\bigg [p_1^2 +m^2 + \sum_i 
(k_{\sigma(i)}^2+2 k_{\sigma(i)} \cdot p_1 )t_i
+ \sum_{i<j} 2k_{\sigma(i)} \cdot k_{\sigma(j)} t_j \bigg]^{-N-1} 
\nonumber \\
&& = N!\int_0^1dt_1  \int_0^{t_1} dt_2 \int_0^{t_2} dt_3 
\cdot\cdot \int_0^{t_{N-1}}  dt_N\, 
\bigg [p_1^2 +m^2 + \sum_i 
\Big [(k_{\sigma(i)}^2+2 k_{\sigma(i)} \cdot (p_1 + \sum_{j=1}^{i-1} 
k_{\sigma(j)})\Big ] t_i \bigg]^{-N-1} 
\nonumber \\
&& = {1\over p_1^2 + m^2}\ 
{1\over (p_1 + k_{\sigma(1)})^2 + m^2}\ 
{1\over (p_1 + k_{\sigma(1)}+ k_{\sigma(2)})^2 + m^2}\, 
\cdot \cdot \cdot \,
{1\over (p_1 + \sum_{i=1}^N k_{\sigma(i)})^2 + m^2}
\eea
This shows that in each internal propagator flows the momentum
as implied by momentum conservation at each vertex.
The last integration above has been carried out by using the well-known formula
\bea
 {1\over A_0 A_1 A_2 \cdot\cdot\cdot A_N} \!\! &=& \!\!
 N!\int_0^1dt_1  \int_0^{t_1} dt_2 \int_0^{t_2} dt_3 
\cdot\cdot \int_0^{t_{N-1}}  dt_N\, 
\\
&\times& \!\!
{1\over [A_0 +(A_1 -A_0)t_1+ (A_2 -A_1)t_2+
\cdot\cdot\cdot + (A_N -A_{N-1})t_N]^{N+1}}
\nonumber 
\eea

\end{appendix}


\end{document}